\documentclass[twocolumn,showpacs,preprintnumbers,unsortedaddress,%
floatfix,amsmath,eqsecnum,aps,pra]{revtex4-1}
\usepackage{bm}% bold math
\usepackage{graphicx}%
\usepackage{calc}%
\usepackage{amsfonts}%
\usepackage{dcolumn}%
\begin{document}

\title{
Exact exchange potential evaluated solely\\
from occupied Kohn-Sham and Hartree-Fock solutions}
\author{ M. Cinal and A. Holas}
\affiliation{Institute of Physical Chemistry of the Polish Academy
of Sciences, ul.\ Kasprzaka 44/52, 01--224 Warszawa, Poland}
\date{\today}
\begin{abstract}
The reported new algorithm determines the exact exchange potential
$v_{\text{x}}$ in a iterative way using energy and orbital shifts
(ES, OS) obtained --- with finite-difference formulas --- from the
solutions (occupied orbitals and their energies) of the
Hartree-Fock-like equation and  the Kohn-Sham-like equation, the
former used for the initial approximation to $v_{\text{x}}$ and the
latter --- for  increments of ES and OS due to subsequent changes of
$v_{\text{x}}$. Thus, solution of the differential equations for OS,
used by K\"ummel and Perdew (KP) [Phys. Rev. Lett. {\bf 90}, 043004
(2003)], is avoided. The iterated  exchange potential, expressed
in terms of ES and OS, is improved  by modifying  ES at odd iteration
steps and OS at even steps. The modification formulas are related to
the OEP equation (satisfied at convergence) written as the condition
of vanishing density shift (DS) --- they are obtained, respectively,
by enforcing its satisfaction  through  corrections to approximate
OS and  by determining optimal ES that minimize the DS norm. The
proposed method, successfully tested for several closed-(sub)shell
atoms, from Be to Kr, within the DFT exchange-only approximation,
proves highly efficient. The calculations using pseudospectral
method for representing orbitals give iterative sequences of
approximate exchange potentials  (starting with the
Krieger-Li-Iafrate approximation) that rapidly approach the exact
$v_{\text{x}}$ so that, for  Ne, Ar and  Zn,
the  corresponding DS norm becomes less than $10^{-6}$
after 13, 13 and  9 iteration steps for a given electron density.
In self-consistent density calculations, orbital energies of
%sub-mHartree
$10^{-4}$ Hartree
accuracy are obtained for these atoms
after, respectively, 9, 12 and 12  density iteration steps,
each involving just 2 steps of $v_{\text{x}}$ iteration, while the
accuracy limit of $10^{-6}$--$10^{-7}$ Hartree is reached after 20
density iterations.

\end{abstract}
\pacs{31.15.E-, 31.15.eg, 31.15.ej}
\maketitle
\section{Introduction}
\label{S1}
%@@@@@@@@@@@@@@@@@@@@@@@@@@@@@@@@@@@@

%The density functional theory (DFT), which is the basis
%of the majority of  modern ab initio calculations,
%
Modern ab initio calculations for molecular and condensed matter systems
are most frequently %mostly
based on the density functional theory (DFT), which
gives the prescription for determining  properties
of an interacting $N$-electron system (in its ground state)
with the aid of the Kohn-Sham (KS) scheme
describing a corresponding system of non-interacting particles
\cite{PY89,DG90,SGWKV05,PK03,HK64}.
While the DFT approach simplifies  enormously the description
of many-electron systems, application of the KS scheme requires the
knowledge of the exchange-correlation (xc) potential which is the part of
the KS Hamiltonian that represents
all complex correlations between electrons in the physical system
(due to their fermionic character  and Coulomb interactions).
The xc potential  is
defined as the functional derivative
$v_{\text{xc}\sigma}({\bm r})=\delta E_{\text
xc}[n_{\uparrow},n_{\downarrow}]/\delta n_{\sigma}({\bm r})$ of the
xc energy $E_{\text{xc}}$ with respect to the density
$n_{\sigma}({\bm r})$ of electrons with spin $\sigma\
(=\uparrow,\downarrow)$ % ---
but its exact form  remains unknown,
so %is unknown,  %and
 it is usually found using the local-density or generalized-gradient
approximations (LDA, GGA) to
the functional $E_{\text{xc}}[n_{\uparrow},n_{\downarrow}]$.
Although this approach has proved to be highly successful in calculations
of numerous physical properties,
it fails in some cases (e.g., for energies of bound unoccupied states)
due to the incomplete cancelation of
Coulomb self-interaction in the KS potential including   $v_{\text{xc}\sigma}$
obtained with LDA or GGA.
%(which leads to, in particular,
%an incorrect large-$r$ dependence of $v_{\text{x}\sigma}({\bm r})$).
Such shortcomings of these approximate xc potentials
are largely removed when
the exact exchange potential
$v_{\text{x}\sigma}({\bm r})$,
which is the dominant part of
$v_{\text{xc}\sigma}=v_{\text{x}\sigma}+v_{\text{c}\sigma}$,
%is  determined exactly.
is used.
It is  defined as $v_{\text{x}\sigma}({\bm r})=
\delta E_{\text{x}}[n_{\sigma}]/\delta n_{\sigma}({\bm r})$
with the $\sigma$-subsystem exchange energy $E_{\text{x}}$
%which is
given explicitly
in terms of the occupied KS orbitals
%$\phi_{a\sigma}$ ($a=1,\ldots,N_{\sigma}$),
which are obtained with
the potential $v_{\text{s}\sigma}$ that corresponds to $n_{\sigma}$
%(this correspondence holds
(by virtue of the Hohenberg-Kohn theorem \cite{HK64}).
The KS potential $v_{\text{s}\sigma}({\bm r})$ including
the so defined exact $v_{\text{x}\sigma}$ is free from self-interaction
%(unlike for $v_{\text{x}\sigma}$ obtained in LDA and GGA)
and, consequently, it has the proper asymptotic dependence ($-1/r$ for atoms),
unlike $v_{\text{s}\sigma}({\bm r})$ obtained with LDA or GGA.
%The exact exchange potential  also shows the characteristic bumps
%due to the atomic-shell structure which  are absent in the LDA and GGA exchange
%(and xc) potentials.

%The exact exchange potential
%$v_{\text{x}\sigma}({\bm r})=
%\delta E_{\text{x}}[n_{\sigma}]/\delta n_{\sigma}({\bm r})$
%is defined with
%the $\sigma$-subsystem exchange energy $E_{\text{x}}$
%given explicitly
%in terms of the occupied KS orbitals
%%$\phi_{a\sigma}$ ($a=1,\ldots,N_{\sigma}$),
% obtained with
%the potential $v_{\text{s}\sigma}$ that corresponds to $n_{\sigma}$
%(the latter correspondence holds by virtue of the Hohenberg-Kohn theorem \cite{HK64}).

%
 The KS orbitals $\phi_{a\sigma}({\bm r})$ and
their energies $\epsilon_{a\sigma}$ are eigensolutions of the KS equation
\begin{subequations}
\label{eq-KS}
\begin{equation}
\hat h_{\text{s}\sigma}({\bm r})\,\phi_{a\sigma}({\bm r}) =
\epsilon_{a\sigma}\,\phi_{a\sigma}({\bm r})\,,\ \ \epsilon_{a\sigma}
< \epsilon_{a+1,\sigma}\,, \label{eq-KS-eqn}
\end{equation}
where
\begin{equation}
\hat h_{\text{s}\sigma}({\bm r}) = \hat t({\bm r}) +
v_{\text{s}\sigma}({\bm r})\,,\ \ \  \hat t({\bm r}) =
-{\textstyle\frac{1}{2}}\bm{\nabla}^2({\bm r}) \label{eq-KS-ham}
\end{equation}
\end{subequations}
(atomic units used throughout the paper, orbitals chosen to be
real) while
the effective local KS potential
$v_{\text{s}\sigma}(\bm r)$ is the sum of  the external,
electrostatic and xc terms.
%For convenience, the KS potential $v_{\text{s}\sigma}$ is modified
This potential is modified, for convenience in  theoretical considerations,
by adding an
infinitesimal symmetry-breaking term to remove any possible degeneracy
(cf. Ref. \onlinecite{KKGG99}),
%It is also convenient for the sake of  theoretical considerations
%to further modify $v_{\text{s}\sigma}$
and by enclosing
%to enclose
the system in a large box
to have fully discrete energy spectrum
(and to guarantee convergence of spatial integrals).
%since it leads to fully discrete energy spectrum and
%guarantees convergence of spatial integrals.
%However, the latter  modification of $v_{\text{s}\sigma}$
The latter  modification of $v_{\text{s}\sigma}$ is not required for bound orbitals
and it is not used by us in practical calculations of %determination
their wavefunctions and energies  with the presently applied
numerical method  (Sec. \ref{S7-num-meth}).
The spin-up, $n_{\uparrow}({\bf r})$, and spin-down,
$n_{\downarrow}({\bf r})$, electron densities
are determined  with  the occupied KS
orbitals $\phi_{a\sigma}$ ($a=1,\ldots,N_{\sigma}$,
$N_{\uparrow}+N_{\downarrow}=N$) through the relation
%  The occupied KS
%orbitals, labeled with $a=1,\ldots,N_{\sigma}$, determine the
%spin-up, $n_{\uparrow}({\bf r})$, and spin-down,
%$n_{\downarrow}({\bf r})$, electron densities through the relation
\begin{equation}
n_{\sigma}({\bm r}) = n[\{\phi_{a\sigma}\}]({\bm r}) \equiv
\sum_{a=1}^{N_\sigma} \phi_{a\sigma}^2({\bm r})\,.
\label{eq-density}
\end{equation}
Since the exchange potential $v_{\text{x}\sigma}$ depends only on
the density $n_{\sigma}$ and it is expressible in terms of
$\sigma$-subsystem characteristics,
the dependence on the spin index $\sigma$
in $v_{\text{x}\sigma}$, $n_{\sigma}$, $N_{\sigma}$,
$\phi_{a\sigma}$ etc. can be suppressed hereafter
and all following discussion refers to one of the
spin-up or spin-down sub-systems separately.

The exact exchange potential  $v_{\text{x}}({\bm r})=
\delta E_{\text{x}}[n]/\delta n({\bm r})$,
defined with exact $E_{\text{x}}$,
satisfies
%leads to
an integral equation the kernel of  which
is the static KS linear response function
involving all (occupied and unoccupied)
KS orbitals of given spin.
This equation was  originally derived within
the optimized-effective-potential (OEP) approach \cite{TSh76,WPC90,KLI92,E03}
and it can be also obtained \cite{GL94} when $v_{\text{x}\sigma}(\bf r)$
is identified as the first-order term in the xc potential expansion
resulting from adiabatic switching the electron-electron interaction
in the many-electron Hamiltonian \cite{GL93}.
The OEP equation for $v_{\text{x}}$
can be  solved numerically
within a spatial grid  representation for atoms
\cite{TSh76,KLI92,LKI93,EV93a,ED99,EHD00}
but the involved determination of the KS response function matrix
makes this approach inefficient for larger systems.
The solution of the OEP equation with
a finite basis for orbitals  and an auxiliary basis for
the  response function \cite{GL94,G99,IHB99}
is usually very troublesome
for atoms and molecules since it
requires careful balancing of the two bases
to avoid  unphysical oscillations
of $v_{\text{x}}({\bm r})$ \cite{HIGBBT01,DSG01,HGSG07}.
Similar problems due to basis imbalance are encountered
\cite{SSD06,HBBY07,HBY08,GHJL08}
when the exact $v_{\text{x}}$, expressed in an auxiliary basis,
is found by following directly the OEP approach, i.e.,
 by minimizing the HF-like energy
expression for total energy given in terms of occupied
(KS) orbitals obtained with a local potential $v_{\text{s}}$;
see Ref. \onlinecite{YW02}.
An alternative method of solving the OEP equation in orbital basis representation, which
avoids the use of  an auxiliary basis  by expressing  potentials
in terms of the products of occupied and unoccupied KS orbitals,
has been proposed very recently \cite{FKF10}.

A different approach to determination of the  exact $v_{\text{x}}$,
introduced by K\"ummel and Perdew (KP) \cite{KP03},
uses the OEP equation rewritten in terms of the occupied KS orbitals and
the corresponding orbital shifts (OS) \cite{KLI92,GKKG00}; see Sec. \ref{S3A}.
The latter are solutions of
$N$ non-homogeneous linear differential equations, hearafter called
the KP equations, which depend on the unknown $v_{\text{x}}$.
The so posed problem of determining $v_{\text{x}}$ (for a given density)
has so far been solved in two ways (Sec.\ \ref{S3C}):
(i) by iterating   $v_{\text{x}}$ and the OS
\cite{KP03},
(ii) with our previously developed algorithm \cite{CH07}
where $v_{\text{x}\sigma}$ is found without any iteration,
using the solution of the system of equations (algebraic and differential)
obtained by a suitable combination of the OEP and KP equations.
A further discussion on
the exact exchange potential
and its orbital-dependent approximations, methods of their determination
and applications
can be found
in the recent  review \cite{KK08} by K\"{u}mmel and  Kronik.

In the present work, we determine the exact  exchange potential
with a novel iterative method based on
the representation of approximate $v_{\text{x}}$
(at each iteration step) in  terms of
the corresponding OS and energy shifts (ES).
Such representation was previously
derived in Refs. \onlinecite{KLI92,GKKG00,KP03}
for  the exact $v_{\text{x}}$, but it is now recognized to be
an identity relation holding for
any local multiplicative potential (Sec. \ref{S2}),
in a similar way as
the relations previously found in Ref.\ \onlinecite{HC05}.
%
%the identity relation between a local multiplicative potential.
%and its OS and energy shifts (ES)
%
%In the present work, we determine the exact  exchange potential
%with a novel method where the approximate $v_{\text{x}\sigma}$
%is expressed, at each iteration step, in terms of the corresponding
%the OS and energy shifts (ES); this identity relation, previously
%derived in Refs. \onlinecite{KLI92,GKKG00,KP03}
%for  the exact $v_{\text{x}\sigma}$, is now recognized to hold for
%any local multiplicative potential (Sec. \ref{S2}).
%
Our new algorithm, summarized in Sec. \ref{S6},
consists of (i) calculating  the OS and ES
with a finite-difference formula
from solutions of the KS(-like) and Hartree-Fock(HF)-like equations
(Sec. \ref{S4}), and
(ii) iterating $v_{\text{x}}$ (for a given density) by
modification of its ES and OS terms with an appropriate use of the OEP equation
(Secs. \ref{S3D}, \ref{S5}),
(iii) iterating the KS potential $v_{\text{s}}$ to obtain the convergent
density.
Thus, solution of the KP equations is avoided, while
the iteration of $v_{\text{x}}$
is done in a different way than in Ref. \onlinecite{KP03}.
The method for calculating of the OS and ES
follows a recent finding that the OS are well represented by the differences of
the respective occupied KS and HF orbitals, see Table I in Ref. \onlinecite{C10}
(note that in the latter work HF orbital-specific exchange
potentials  are shown to map very accurately onto  $v_{\text{x}}$
%which explains
and thus
the outstanding proximity
between the HF and KS orbitals
is explained).
%
%(therein, HF orbital-specific exchange
%potentials are shown to map very accurately onto  $v_{\text{x}}$
%which explains the outstanding proximity
%between the HF and KS orbitals).
%

The efficiency of the proposed algorithm is tested
(Sec. \ref{S7}) for several closed-(sub)shell atoms
from Be to Kr within the exchange-only KS scheme where the correlation
energy $E_\text{c}$ is neglected together with the potential
$v_{\text{c}}$, a part of the xc potential, thus leaving  only
$v_{\text{x}}$.
In the performed tests, the obtained exchange
potentials, which are, in fact, very accurate approximations
to the exact $v_{\text{x}}$,
are compared with
the exact exchange potential (used as a reference)
determined with extremely high accuracy using our non-iterative algorithm \cite{CH07}.
Similar comparison is done between the corresponding
electron densities and orbital energies.
\section{ Identity satisfied by any local potential}
\label{S2}
%@@@@@@@@@@@@@@@@@@@@@@@@@@@@@@@@@@@@@@

We find it useful to begin our presentation of a new algorithm for
the exact exchange potential $v_{\text{x}}({\bm r})\equiv
v_{\text{x}}^{\text{E}}({\bm r})$ by our
reformulation of the investigation of Krieger {\em et al.\ }%
\cite{KLI92}, continued next by Grabo {\em et al.\ }\cite{GKKG00}
and by K\"ummel and Perdew \cite{KP03}. It is given in a form of an
identity satisfied by an arbitrary local (multiplicative) potential
$v_{\text{x}}^{\text{L}}({\bm r})$ [some restriction on it will be
indicated below Eq.\ (\ref{eq-OS-constr})]. This identity is written
in terms of the occupied KS orbitals $\{\phi_a({\bm r})\}_{a=1}^N$,
Eq.\ (\ref{eq-KS-eqn}) [directly and via the density $n({\bm r})$,
Eq.\ (\ref{eq-density})], auxiliary constants
--- the energy shifts (ES) $\{D_{aa}^{\text{L}}\}_{a=1}^N$ and
auxiliary functions
--- the orbital shifts (OS) $\{\delta\phi_a^{\text{L}}({\bm
r})\}_{a=1}^N$, namely
\begin{eqnarray}
v_{\text{x}}^{\text{L}}({\bm r})&=&
v_{\text{x}}^{\text{Sl}}[\{\phi_a\}]({\bm r})+
v_{\text{x}}^{\text{ES}}[\{D_{aa}^{\text{L}}\},\{\phi_a\}]({\bm r})
\nonumber\\
&& \mbox{}+
v_{\text{x}}^{\text{OS}}[\{\delta\phi_a^{\text{L}}\},\{\phi_a\}]({\bm
r})\,, \label{eq-identity}
\end{eqnarray}
where
\begin{equation}
v_{\text{x}}^{\text{Sl}}[\{\phi_a\}]({\bm r})= \frac {1}{n({\bm r})}
\sum_{a=1}^{N} \phi_a({\bm r})
\hat{v}_{\text{x}}^{\text{F}}[\{\phi_b\}]({\bm r})\phi_a({\bm r})
\label{eq-vx-Slater}
\end{equation}
is known as the Slater potential, while
\begin{subequations}
\label{eq-vx-ES}
\begin{equation}
v_{\text{x}}^{\text{ES}}[\{D_{aa}^{\text{L}}\},\{\phi_a\}]({\bm r})
\equiv v_{\text{x}}^{\text{ESL}}({\bm r}) = \sum_{a=1}^{N}
v_{\text{x}}^{aa}({\bm r}) D_{aa}^{\text{L}}\,,\label{eq-vx-ESL}
\end{equation}
where
\begin{equation} v_{\text{x}}^{aa}({\bm r}) =
\frac{\phi_a^2({\bm r})}{n({\bm r})}\,, \label{eq-vx^aa}
\end{equation}
\end{subequations}
and
$$v_{\text{x}}^{\text{OS}}[\{\delta\phi_a^{\text{L}}\},\{\phi_a\}]({\bm
r}) \equiv v_{\text{x}}^{\text{OSL}}({\bm r}) = \sum_{a=1}^{N}\hat
v_{\text{x}a}^{\text{OS}}({\bm r})\, \delta\phi_a^{\text{L}}({\bm
r})\equiv$$
\begin{equation}
\frac {1}{n({\bm r})}\sum_{a=1}^{N}
\biglb\{\delta\phi_a^{\text{L}}({\bm r})\,\hat t({\bm r})\phi_a({\bm
r}) - \phi_a({\bm r})\,\hat t({\bm r})\delta\phi_a^{\text{L}}({\bm
r}) \bigrb\}\,, \label{eq-vx-OS}
\end{equation}
are the ES and OS components of the potential. The OS operator $\hat
v_{\text{x}a}^{\text{OS}}$ [occurring in Eq.\ (\ref{eq-vx-OS})]
includes a differential operator $\hat t({\bm r})$, while the Fock
exchange operator $\hat{v}_{\text{x}}^{\text{F}}$ [occurring in Eq.\
(\ref{eq-vx-Slater})] is a non-local integral one defined by
\begin{equation}
\hat{v}_{\text{x}}^{\text{F}}[\{\phi_a\}]({\bm r})\,\varphi({\bm
r})= - \int \text{d}^3r'\sum_{a=1}^N \frac{\phi_a({\bm
r})\,\phi_a({\bm r}')\,\varphi({\bm r}')}{|{\bm r}-{\bm r}'|}\,.
\label{eq-Fock-op}
\end{equation}
The ES and OS for the potential $v_{\text{x}}^{\text{L}}$ are given
by
\begin{eqnarray}
D_{aa}^{\text{L}} &=&
D_{aa}[v_{\text{x}}^{\text{L}}-\hat{v}_{\text{x}}^{\text{F}}]\,,
\label{eq-ES-L}\\
\delta\phi_a^{\text{L}}({\bm r}) &=&
\delta\phi_a[v_{\text{x}}^{\text{L}}-\hat{v}_{\text{x}}^{\text{F}}]({\bm
r})\,, \label{eq-OS-L}
\end{eqnarray}
where for any potentials difference $\delta\hat v({\bm r})$ the ES
$D_{aa}[\delta\hat v]$ are defined as its matrix elements
\begin{equation}
D_{aa}[\delta\hat v] = \langle \phi_a|\delta\hat v|\phi_a\rangle\,,
\label{eq-ES-def}
\end{equation}
while the OS $\delta\phi_a[\delta\hat v]({\bm r})$ are solutions of
the differential equation involving $\delta\hat v({\bm r})$
\begin{subequations}
\label{eq-OS-def}
\begin{equation} \Bigl[ \hat t({\bm r}) +
{v}_{\text{s}}({\bm r})-\epsilon_a \Bigr]\delta \phi_a[\delta\hat
v]({\bm r}) =-\hat{w}_a[\delta\hat v]({\bm r})\,\phi_a({\bm r})\,,
\label{eq-OS-diff}
\end{equation}
where the right-hand side operator is
\begin{equation}
\hat{w}_a[\delta\hat v]({\bm r}) =\delta\hat v({\bm
r})-\langle\phi_a|\delta\hat v| \phi_a\rangle\,. \label{eq-OS-w}
\end{equation}
These solutions must satisfy the same boundary conditions as
orbitals. In addition, they must satisfy the orthogonality
requirement
\begin{equation}
\langle \phi_a|\delta\phi_a[\delta\hat v]\rangle = 0\,.
\label{eq-OS-constr}
\end{equation}
The identity (\ref{eq-identity}) concerns only such potentials
$v_{\text{x}}^{\text{L}}$ that the solution
$\delta\phi_a[v_{\text{x}}^{\text{L}}-
\hat{v}_{\text{x}}^{\text{F}}]$ of Eq.\ (\ref{eq-OS-def}) exists.
Note that $\langle\phi_a|\hat{w}_a|\phi_a\rangle=0$, and that
$\hat{w}_a[\delta\hat v]$ is invariant with respect to a constant
shift $c$ of its argument, therefore
\begin{equation}
\delta\phi_a[\delta\hat v + c]({\bm r}) = \delta\phi_a[\delta\hat v
]({\bm r})\,. \label{eq-OS-shift}
\end{equation}
\end{subequations}

As we see, all terms on the right-hand side of the identity
(\ref{eq-identity}) can be evaluated knowing only
$v_{\text{x}}^{\text{L}}$, the occupied KS solutions $\{\phi_a,
\epsilon_a\}_{a=1}^N$ and the KS potential ${v}_{\text{s}}$. The
dependence on $v_{\text{x}}^{\text{L}}$ enters via the functionals
ES and OS, Eqs.\ (\ref{eq-ES-L}) and (\ref{eq-OS-L}). The dependence
on ${\bm r}$ enters the separate components  via $\{\phi_a({\bm
r})\}$, $n({\bm r})$, $\hat{v}_{\text{x}}^{\text{F}}({\bm r})$ and
$\{\delta\phi_a^{\text{L}}({\bm r})\}$. The identity
(\ref{eq-identity}) is similar to other three identities obtained by
us in our paper \cite{HC05}. Any local potential potential expressed
according to the the identity (\ref{eq-identity}) will be named to
have the {\em canonical} form. This identity will be used by us for
$v_{\text{x}}^{\text{L}}$ representing an approximate exchange
potential in the process of iterative improvement leading to the
exact one.

The validity of Eq.\ (\ref{eq-identity}) can be proven quite simply.
For $\delta\hat v =v_{\text{x}}^{\text{L}} -
\hat{v}_{\text{x}}^{\text{F}}$ one should multiply both sides of
Eq.\ (\ref{eq-OS-diff}) by $\phi_a({\bm r})$, sum up over $a$ from 1
to $N$, use definitions (\ref{eq-density}) and (\ref{eq-KS-ham}),
and utilize  Eq.\ (\ref{eq-KS-eqn}) to replace $\epsilon_a\,\phi_a$
by the equivalent expression. Locality of $v_{\text{x}}^{\text{L}}$
and $v_{\text{s}}$ is to be taken into account. It should be noted
that the requirement (\ref{eq-OS-constr}) is not involved in this
procedure. However, it defines uniquely the particular solution of
Eq.\ (\ref{eq-OS-diff}) that allows for expressing
$\delta\phi_a^{\text{L}}$ in the equivalent perturbation-theory
form, Sec.\ \ref{S4}. A general solution is $\delta\phi_a^{\text{L}}
+ c_a \phi_a$, with an arbitrary constant $c_a$.

There are interesting properties of the canonical representation of
a local potential. The replacement
$\delta\phi_a^{\text{L}}\rightarrow \delta\phi_a^{\text{L}} + c_a
\phi_a$ (with $c_a$ --- a constant) does not change the OS
potential, Eq.\ (\ref{eq-vx-OS}). While the OS, Eq.\
(\ref{eq-OS-L}), are insensitive to the replacement
$v_{\text{x}}^{\text{L}}\rightarrow v_{\text{x}}^{\text{L}} +
c_{\text{x}}^{\text{L}}$ (with $c_{\text{x}}^{\text{L}}$ --- a
constant), see Eq.\ (\ref{eq-OS-shift}), the ES, Eq.\
(\ref{eq-ES-L}), are changed by this replacement
\begin{equation}
D_{aa}^{\text{L}}\rightarrow
D_{aa}^{\text{L}}+c_{\text{x}}^{\text{L}}\,. \label{eq-ES-shift}
\end{equation}
Therefore, due to the obvious identity [see Eq.\ (\ref{eq-vx^aa})]
\begin{equation}
\forall {\bm r} \,,\ \ \sum_{a=1}^{N} v_{\text{x}}^{aa}({\bm r}) =
1\,,\label{eq-v_aa-ident}
\end{equation}
the ES potential, Eq.\ (\ref{eq-vx-ESL}), transforms as
\begin{equation}
v_{\text{x}}^{\text{ES}}({\bm r})\rightarrow
v_{\text{x}}^{\text{ES}}({\bm r})+c_{\text{x}}^{\text{L}}\,.
\label{eq-vxES-shift}
\end{equation}
Thus, the identity (\ref{eq-identity}) remains valid for the
potential $v_{\text{x}}^{\text{L}}$ replaced by
$(v_{\text{x}}^{\text{L}} + c_{\text{x}}^{\text{L}})$ on both sides
of the identity.

It proves convenient for further applications to fix the additive
constant of an approximate exchange potential
$v_{\text{x}}^{\text{L}}({\bm r})$ by the replacement
\begin{equation}
D_{aa}^{\text{L}}:=D_{aa}^{\text{L}}-D_{NN}^{\text{L}}\,,
\label{eq-ES-repl}
\end{equation}
resulting in
\begin{equation}
D_{NN}^{\text{L}} = 0\,. \label{eq-D_NN0}
\end{equation}

With Eq.\ (\ref{eq-D_NN0}) satisfied, the potential
$v_{\text{x}}^{\text{ES}}[\{D_{aa}^{\text{L}}\},\{\phi_a\}]({\bm
r})$ is exponentially small in the large-$r$ region for a {\em
general} direction ${\bm r}/r$ (i.e. for ${\bm r}$ not laying on the
nodal surface of $\phi_N({\bm r})$, if this orbital possesses such
surface extending to the asymptotic region, see Della Sala and
G\"{o}rling, \cite{DSG02}). The Slater potential
$v_{\text{x}}^{\text{Sl}}({\bm r})$ is known to behave
asymptotically as $-1/r$. Thus, the asymptotic behavior of the OS
potential is needed to find the asymptotic form of
$v_{\text{x}}^{\text{L}}({\bm r})$ from its canonical
representation.

For the exact (E) exchange potential, $v_{\text{x}}^{\text{L}}
=v_{\text{x}}^{\text{E}} $, its OS part
$v_{\text{x}}^{\text{OSE}}({\bm r})$, Eq.\ (\ref{eq-vx-OS}),
exponentially vanishes at $r\rightarrow\infty $, so it is
asymptotically smaller than $v_{\text{x}}^{\text{Sl}}({\bm r})$.
This is a consequence of the OEP equation [see Eq.\ (\ref{eq-OEP})
below] which implies the OS $\delta\phi_{N-1}^{\text{E}}$ and
$\delta\phi_N^{\text{E}}$ in the large-$r$ region must behave (with
accuracy to their leading exponential factors) as $-C_N\phi_N$ and
$C_N\phi_{N-1}$, respectively, in order to satisfy $\sum_{a=1}^N
2\phi_a\, \delta\phi_a^{\text{E}}=0$. Then
$v_{\text{x}}^{\text{OSE}}$ will be dominated by the terms with
$a=N-1$ and $a=N$, each as small as $\phi_{N-1}/\phi_N$.

However,  for an arbitrary local potential
$v_{\text{x}}^{\text{L}}$, which does not satisfy the OEP equation,
the respective OS potential $v_{\text{x}}^{\text{OSL}}({\bm r})$ may
be finite at   $r \rightarrow \infty$. In this case, imposing the
condition (\ref{eq-D_NN0}) --- though it results in the
asymptotically vanishing term $v_{\text{x}}^{\text{ESL}}$ (for a
{\em general} direction) --- leads to $v_{\text{x}}^{\text{L}}({\bm
r})$ that tends to a finite value $v_{\text{x}}^{\text{L}}(\infty) =
v_{\text{x}}^{\text{OSL}}(\infty)$.
\section{ Exact exchange potential}
\label{S3}
%@@@@@@@@@@@@@@@@@@@@@@@@@@@@@@@
%
\subsection{Density shift and the OEP equation}
\label{S3A}
%&&&&&&&&&&&&&&&&&&&&
Considering the exchange potential to be a functional of the given
density of the KS system, we will suppress often the dependence of
various objects on the fixed KS characteristics
$\{\phi_b,\epsilon_b\},v_{\text{s}}$. As a tool to define the exact
exchange potential, we introduce the (first-order in OS) density
shift (DS) induced by $\phi_a \rightarrow \phi_a +
\delta\phi_a^{\text{L}}$ in Eq.\ (\ref{eq-density})
\begin{eqnarray}
\delta n^{\text{L}}({\bm r}) &\equiv& \delta n
[\{\delta\phi_b^{\text{L}}\}]({\bm r}) = \sum_{a=1}^{N} 2\,
\phi_a({\bm r})\,\delta\phi_a^{\text{L}}({\bm r})
\label{eq-den-shift}\\
&\equiv&\delta\breve n[v_{\text{x}}^{\text{L}}]({\bm r}) =
\sum_{a=1}^{N} 2\,\phi_a({\bm
r})\,\delta\phi_a[v_{\text{x}}^{\text{L}}-
\hat{v}_{\text{x}}^{\text{F}}]({\bm r})\,, \nonumber
\end{eqnarray}
where the OS $\delta\phi_a$ are defined in Eqs.\ (\ref{eq-OS-L}) and
(\ref{eq-OS-def}).

As well known, the exact (E) exchange potential, to be denoted here
$v_{\text{x}}({\bm r})\equiv v_{\text{x}}^{\text{E}}({\bm r})$, can
be found as the solution of the so called ``optimized effective
potential'' (OEP) equation \cite{KLI92,GKKG00,KP03}
\begin{subequations}
\label{eq-vx^E-def}
\begin{equation}
\forall {\bm r}\,,\ \ \ \delta\breve n
[v_{\text{x}}^{\text{E}}]({\bm r}) = 0\,.\label{eq-OEP}
\end{equation}
But this solution can be determined only with accuracy to an
additive constant. Therefore, to make it unique, the additional
requirement on $v_{\text{x}}^{\text{E}}$ is imposed
\begin{equation}
\lim_{ r\rightarrow \infty} v_{\text{x}}^{\text{E}}({\bm r}) = 0\,,\
\ \text{for a general direction\ }{\bm r}/r\,.\label{eq-vx^E-asym}
\end{equation}
\end{subequations}
Then $D_{NN}^{\text{E}}=0$ is satisfied.
\subsection{Singular components of OS potential}
\label{S3B}
%&&&&&&&&&&&&&&&&&&&&
Since the identity (\ref{eq-identity}) will be helpful in finding
the solution of the OEP Eq.\ (\ref{eq-OEP}), we make use of the DS
$\delta n({\bm r})$ to rewrite the OS potential, Eq.\
(\ref{eq-vx-OS}), in two forms:
\begin{subequations}
\label{eq-vx-OS-forms}
\begin{eqnarray}
&&\hspace{-2em}v_{\text{x}}^{\text{OSa}}[\{\delta\phi_a^{\text{L}}\},\delta
n^{\text{a}}]({\bm r})
\nonumber\\
&=& \frac{1}{n({\bm r})}\Biggl(\,\sum_{a=1}^{N} \biglb\{ -
\bm{\nabla} \bigl(\delta\phi_a^{\text{L}}({\bm
r})\bm{\nabla}\phi_a({\bm
r})\bigr)\bigrb\} \nonumber\\
&&\hspace{3em}\mbox{} -{\textstyle\frac{1}{2}}\,\hat t\,\delta
n^{\text{a}}({\bm r})\Biggr)\,,
\label{eq-vx-OS-fa}\\
&&\hspace{-2em}v_{\text{x}}^{\text{OSb}}[\{\delta\phi_a^{\text{L}}\},\delta
n^{\text{b}}]({\bm r})
\nonumber\\
&=&\frac{1}{n({\bm r})}\Biggl(\,\sum_{a=1}^{N} \biglb\{2 \epsilon_a
\phi_a({\bm r}) - \bigl(\bm{\nabla}\phi_a({\bm r})\bigr)
\!\cdot\!\bm{\nabla}\bigrb\} \delta\phi_a^{\text{L}}({\bm r}) \nonumber\\
&&\hspace{3em}\mbox{}-\bigl(v_{\text{s}}({\bm
r})+{\textstyle\frac{1}{2}}\,\hat t\,\bigr)\,\delta
n^{\text{b}}({\bm r})\Biggr)\,, \label{eq-vx-OS-fb}
\end{eqnarray}
equivalent to the original OS potential only when $\delta
n^{\text{a}}$ and $\delta n^{\text{b}}$ are appropriate [i.e., given
by Eq.\ (\ref{eq-den-shift})]
\begin{eqnarray}
\!\!\!\!\!\!v_{\text{x}}^{\text{OS}}[\{\delta\phi_a^{\text{L}}\}]({\bm
r})&\equiv&
v_{\text{x}}^{\text{OSa}}\Bigl[\{\delta\phi_a^{\text{L}}\},\delta n
\bigl[\{\delta\phi_a^{\text{L}}\}\bigr]\Bigr]({\bm r})\nonumber\\
&\equiv&
v_{\text{x}}^{\text{OSb}}[\{\delta\phi_a^{\text{L}}\},\delta n
\bigl[\{\delta\phi_a^{\text{L}}\}\bigr]\Bigr]({\bm r}).
\label{eq-vx-OS-fc}
\end{eqnarray}
\end{subequations}

For any $v_{\text{x}}^{\text{L}}({\bm r})$ represented by the
identity (\ref{eq-identity}), finite everywhere, three components
$v_{\text{x}}^{\text{Sl}}({\bm r})$,
$v_{\text{x}}^{\text{ES}}[\{D_{aa}^{\text{L}}\}]({\bm r})$,
$v_{\text{x}}^{\text{OS}}[\{\delta\phi_a^{\text{L}}\}]({\bm r})$ are
also finite when calculated with the KS characteristics. But this
may be not true for separate terms of $v_{\text{x}}^{\text{OS}}$
represented in Eq.\ (\ref{eq-vx-OS-forms}). Coulombic singularity of
the KS potential at each nucleus position results in kinks of KS
orbitals and density (Kato cusp), and also in kinks of
$\delta\phi_a^{\text{L}}({\bm r})$ [when determined from Eqs.\
(\ref{eq-OS-L}) and (\ref{eq-OS-def})] and, therefore, in kinks of
$\delta n[\{\delta\phi_a^{\text{L}}\}]({\bm r})$. It can be easily
verified by expansions (in displacements from the point of
singularity) that $v_{\text{s}}({\bm r})\,\delta
n[\{\delta\phi_a^{\text{L}}\}]({\bm r})$ and
${\textstyle\frac{1}{2}}\,\hat t({\bm r})\,\delta
n[\{\delta\phi_a^{\text{L}}\}]({\bm r})$ taken separately are
singular, while their sum $\bigl(v_{\text{s}}({\bm
r})+{\textstyle\frac{1}{2}}\,\hat t({\bm r})\,\bigr)\,\delta
n[\{\delta\phi_a^{\text{L}}\}]({\bm r})$ remains finite due to
cancelation of singularities. Therefore the finite potentials
$v_{\text{x}}^{\text{OSa}}$ and $v_{\text{x}}^{\text{OSb}}$ with
arguments indicated in Eq.\ (\ref{eq-vx-OS-fc}) behave differently
after truncation by means of inserting $\delta n\equiv 0$. The
truncated OS potential $v_{\text{x}}^{\text{OSa}}[\{\delta
\phi_a^{\text{L}}\},0\,]({\bm r})$ is singular at nucleus position
(because the neglected term was singular), while
$v_{\text{x}}^{\text{OSb}}[\{\delta\phi_a^{\text{L}}\},0\,]({\bm
r})$ is finite (because the neglected term was finite).
\subsection{Two known approaches to solve OEP equation in terms of orbital shifts}
\label{S3C}
%&&&&&&&&&&&&&&&&&&&&
There are known two approaches to solve the OEP Eq.\ (\ref{eq-OEP})
for $v_{\text{x}}^{\text{E}}({\bm r})$ by evaluating the OS as
auxiliary functions: the iterative procedure, and the one-step
procedure. In the last case, conceived by us and described in Ref.\
\onlinecite{CH07}, one solves the following system of $(2N+1)$ equations:
(i) one functional Eq.\ (\ref{eq-OEP}) [with Eq.\
(\ref{eq-den-shift}) inserted], (ii) one functional equation [Eq.\
(\ref{eq-identity}) for L=E]
\begin{equation}
v_{\text{x}}^{\text{E}}({\bm r}) = v_{\text{x}}^{\text{Sl}}({\bm
r})+ v_{\text{x}}^{\text{ES}}[\{D_{aa}^{\text{E}}\}]({\bm r}) +
v_{\text{x}}^{\text{OS}}[\{\delta\phi_a^{\text{E}}\}]({\bm r})\,,
\label{eq-vx-E}
\end{equation}
(iii) $(N-1)$ functional differential Eqs.\ (\ref{eq-OS-diff}), (iv)
$(N-1)$ scalar Eqs.\ (\ref{eq-OS-constr}) [in (iii) and (iv), for
$a=2,3,\ldots,N$, and for L=E], (v) one scalar equation
$D_{NN}^{\text{E}}=0$. [Since Eq.\ (\ref{eq-OEP}) is included, the
OS potential  in the form
$v_{\text{x}}^{\text{OSb}}[\{\delta\phi_a^{\text{E}}\},0\,]$, Eq.\
(\ref{eq-vx-OS-fb}), is the most convenient.] The enumerated system
of equations can be solved uniquely for one function
$v_{\text{x}}^{\text{E}}({\bm r})$, for $N$ functions
$\delta\phi_a^{\text{E}}({\bm r}),\,a=1,2,\ldots,N$, and for  $N$
constants $D_{aa}^{\text{E}},\,a=1,2,\ldots,N$. These constants
satisfy Eq.\ (\ref{eq-ES-L}).

The iterative approach was devised, in fact, earlier than ours,
by K\"ummel and Perdew (KP) \cite{KP03}. The improved approximate
exchange potential $v_{\text{x}}^{\text{new}}({\bm r})$ is
calculated from the current approximate exchange potential
$v_{\text{x}}^{\text{old}}({\bm r})$ according to the equation
\begin{equation}
v_{\text{x}}^{\text{new}}({\bm r}) = v_{\text{x}}^{\text{Sl}}({\bm
r})+ v_{\text{x}}^{\text{ES}}[\{D_{aa}^{\text{old}}\}]({\bm r}) +
v_{\text{x}}^{\text{OSa}}[\{\delta\phi_a^{\text{old}}\},0\,]({\bm
r}) \label{eq-vx-KP-recur1}
\end{equation}
(with $D_{NN}^{\text{old}}=0$ imposed). This equation follows from
the identity (\ref{eq-identity}) (valid for arbitrary local
potential) ``spoiled'' by inserting the target relation $\delta
n^{\text{a}}({\bm r})=\delta\breve n({\bm r})=0$
--- the OEP Eq.\ (\ref{eq-OEP}) for the exact exchange potential ---
into the version $v_{\text{x}}^{\text{OSa}}$ of the OS potential. As
follows from our analysis given below Eq.\ (\ref{eq-vx-OS-fc}), the
OS term in Eq.\ (\ref{eq-vx-KP-recur1}) calculated for
$v_{\text{x}}^{\text{old}}({\bm r})\ne v_{\text{x}}^{\text{E}}({\bm
r})$ (i.e., before convergence) is divergent at nucleus position. KP
do not comment this feature. According to KP, the iteration cycle
based on Eq.\ (\ref{eq-vx-KP-recur1}) converged quite well to the
exact potential for extended electronic systems, but there were
numerical difficulties with evaluating the OS potential term for
finite systems in the large-$r$ region. Therefore alternative
recurrence relation was proposed and applied by KP
\begin{equation}
v_{\text{x}}^{\text{new}}({\bm r}) = v_{\text{x}}^{\text{old}}({\bm
r}) + c\,\delta n \bigl[\{\delta\phi_a[v_{\text{x}}^{\text{old}}
-\hat{v}_{\text{x}}^{\text{F}}]\}\bigr]({\bm r})\,,
\label{eq-KP-recur2}
\end{equation}
(here a constant $c>0$ is a system dependent parameter), see Eq.\
(\ref{eq-den-shift}) for the density shift definition. At
convergence, the second term vanishes, which indicates that the
obtained potential satisfies the OEP Eq.\ (\ref{eq-OEP}). In each
step of the cycle, the OS are calculated by solving a system of
differential equations, see Eq.\ (\ref{eq-OS-def}).
\subsection{New approach proposed}
\label{S3D}
%&&&&&&&&&&&&&&&&&&&&
Proposed by us in the present paper method of determining the exact
exchange potential is also iterative and makes use of the identity
(\ref{eq-identity}). Three essential novelties are introduced: (i)
calculation of the OS that avoids solving the specific differential
equation --- it will be described in Sec.\ \ref{S4}, (ii)
variational determination of the optimal ES described in Sec.\
\ref{S5}, and (iii) a new recurrence relation for improving
potential, where the OS potential is evaluated with the help of the
modified OS vector
$(\delta\phi_1^{\text{L}\perp},\delta\phi_2^{\text{L}\perp},
\ldots,\delta\phi_N^{\text{L}\perp})$
--- the component of the original OS vector  $(\delta\phi_1^{\text{L}},
\delta\phi_2^{\text{L}}, \ldots,\delta\phi_N^{\text{L}})$
perpendicular to the orbitals vector $(\phi_1,\phi_2,
\ldots,\phi_N)$, namely, $\forall{\bm r}$,
\begin{equation}
\delta\phi_a^{\text{L}\perp}({\bm r}) = \delta\phi_a^{\text{L}}({\bm
r}) -  \phi_{a}({\bm r})\,\delta n[\{\delta\phi_b^{\text{L}}\}]({\bm
r})/\bigl(2 n({\bm r})\bigr)\,. \label{eq-OSperp}
\end{equation}
By construction, the modified OS satisfy the OEP equation
\begin{equation}
\forall {\bm r}\,,\ \ \ \delta n
[\{\delta\phi_b^{\text{L}\perp}\}]({\bm r}) = \sum_{a=1}^{N} 2
\phi_a({\bm r})\,\delta\phi_a^{\text{L}\perp}({\bm r}) = 0\,.
\label{eq-OSperp-OEP}
\end{equation}
Therefore, at convergence (L=E), the second term in Eq.\
(\ref{eq-OSperp}), i.e., the modification to
$\delta\phi_a^{\text{L}}$, vanishes. Of course, before the
convergence is achieved, $\delta\phi_a^{\text{L}\perp}$ does not
satisfy Eq.\ (\ref{eq-OS-L}), therefore the identity
(\ref{eq-identity}) is ``spoiled'' by using
$\delta\phi_a^{\text{L}\perp}$ instead of $\delta\phi_a^{\text{L}}$.
The proposed by us recurrence relation
\begin{equation}
v_{\text{x}}^{\text{new}}({\bm r}) = v_{\text{x}}^{\text{Sl}}({\bm
r})+ v_{\text{x}}^{\text{ES}}[\{D_{aa}^{\text{old}}\}]({\bm r}) +
v_{\text{x}}^{\text{OS}}[\{\delta\phi_a^{\text{old}\perp}\}]({\bm
r}) \label{eq-vx-CH-recur}
\end{equation}
converges  fast but only when it is used in combination with another
step in the recurrence formula --- the mentioned optimization of ES;
otherwise the convergence is very slow or the iterated
$v_{\text{x}}$ diverges. The relation (\ref{eq-vx-CH-recur})
modifying  the OS term is a necessary ingredient in our approach
since the potential $v_{\text{x}}$ with optimized ES cannot be
further improved by mere repeating the ES optimization (see Sec.
\ref{S5B}).

Due to the satisfaction of Eq.\ (\ref{eq-OSperp-OEP}),
the OS potential entering the expression (\ref{eq-vx-CH-recur})
for $v_{\text{x}}^{\text{new}}$ can be evaluated more conveniently as
$v_{\text{x}}^{\text{OSb}}[\{\delta\phi_a^{\text{old}\perp}\}\,,0\,]$,
Eq.\ (\ref{eq-vx-OS-fb}), the equivalent form, according to Eq.\
(\ref{eq-vx-OS-fc}). There are two main advantages of using the
modified OS: (i) the OEP equation is taken into account in the
original (full) OS potential expression at each step, and (ii) this
modification reduces numerical difficulties (faced by KP) with
evaluating the OS potential term.
\section{ Orbital shifts and energy shifts by the new method}
\label{S4}
%@@@@@@@@@@@@@@@@@@@@@@@@@@@@@@@@
It can be easily checked (see, e.g., in \cite{GKKG00}) that the
definition of the OS given in Eq.\ (\ref{eq-OS-def}) is equivalent
to
\begin{subequations}
\label{eq-PT}
\begin{equation}
\delta\phi_a[\delta \hat v]({\bm r}) = \sum_{j=1,\,j \neq
a}^{\infty} \phi_j({\bf r})\,c_{ja}[\delta \hat v]\,,
\label{eq-OS-PT}
\end{equation}
where
\begin{eqnarray}
 c_{ja}[\delta \hat v] & = & -\frac{D_{ja}[\delta \hat v] }{\epsilon_j -
\epsilon_a}\,, \label{eq-cja} \\
D_{ja}[\delta \hat v] & = & \langle\phi_j|\delta \hat
v|\phi_a\rangle\,. \label{eq-Dja}
\end{eqnarray}
\end{subequations}
It is worth noting that the ES defined in Eq.\ (\ref{eq-ES-def})
agree with the definition (\ref{eq-Dja}) of $D_{ja}[\delta \hat v]$.

But the expressions given in Eqs.\ (\ref{eq-PT}) for
\{OS\,,\,ES\}$\equiv\{\delta \phi_a[\delta \hat v]({\bm r}),
D_{aa}[\delta \hat v]\}$ can be interpreted as the first-order
perturbation-theory corrections to the KS solutions $\{\phi_a({\bm
r}),\epsilon_a\}$, stemming from the  perturbation $\delta \hat
v({\bm r})$ of the KS Hamiltonian $\hat{h}_{\text{s}}({\bm r})$. So
they can be evaluated also as the first derivatives, with respect to
the perturbation-strength (real) parameter $\lambda$, of the
solutions $\{\phi_a^\lambda[\delta \hat v]({\bm
r}),\epsilon_a^\lambda[\delta \hat v]\}$ of the following
one-particle Schr\"odinger equation --- the perturbed KS Eq.\
(\ref{eq-KS-eqn}):
\begin{equation}
\{\hat h_{\text{s}}({\bm r}) +\lambda\,\delta \hat v({\bm r})
\}\,\phi_a^\lambda[\delta \hat v]({\bm r}) =
\epsilon_a^\lambda[\delta \hat v]\,\phi_a^\lambda[\delta \hat
v]({\bm r})\,,\ \epsilon_a^\lambda< \epsilon_{a+1}^\lambda\,,
\label{eq-HF-eqn}
\end{equation}
namely
\begin{eqnarray}
\hspace{-5pt}\delta\phi_a[\delta \hat v]({\bm r})
&\hspace{-3pt}=&\hspace{-3pt} \left.\bigl(\partial
\phi_a^\lambda[\delta \hat v]({\bm r})/
\partial \lambda\bigr)\right|_{\lambda =0}\equiv
\overset\circ\phi_a[\delta
\hat v]({\bm r})\,,\label{eq-PT-der-OS}\\
D_{aa}[\delta \hat v]&\hspace{-3pt}=&\hspace{-3pt}
\left.\bigl(\partial\epsilon_a^\lambda[\delta \hat v] / \partial
\lambda\bigr)\right|_{\lambda =0}\equiv
\overset\circ\epsilon_a[\delta \hat v]\,. \label{eq-PT-der-ES}
\end{eqnarray}
The crucial for our algorithm Eq.\ (\ref{eq-HF-eqn}) for $\delta
\hat v = v_{\text{x}}^{\text{L}}-\hat{v}_{\text{x}}^{\text{F}}$ will
be named the {\em HF-like equation}, because  its Hamiltonian can be
viewed as a modified HF Hamiltonian:
\begin{eqnarray}
&&\hat h_{\text{s}}({\bm r}) +
\lambda\,\bigl(v_{\text{x}}^{\text{L}}({\bm r})
-\hat{v}_{\text{x}}^{\text{F}}({\bm r})\bigr)\nonumber\\
&&= \hat t({\bm r}) + \bigl(v_{\text{s}}({\bm r})+\lambda\,
v_{\text{x}}^{\text{L}}({\bm r})\bigr) - \lambda\,
\hat{v}_{\text{x}}^{\text{F}}({\bm r})\,, \label{eq-HF-ham}
\end{eqnarray}
containing, besides a local potential, the (scaled by $-\lambda$)
characteristic Fock exchange potential operator, Eq.\
(\ref{eq-Fock-op}). Therefore, after minor adaptation, standard HF
codes are suitable for numerical solving Eq.\ (\ref{eq-HF-eqn}) at
any finite $\lambda$. We will also calculate the OS and ES generated
by some local perturbing potential $\delta \bar{v}({\bm r})$ instead
of nonlocal $\delta\hat v({\bm r})$. The corresponding perturbed KS
equation [like Eq.\ (\ref{eq-HF-eqn}), but with $\delta\hat v$
replaced by $\delta \bar v$] will be named the {\em KS-like
equation}.

Having solutions of Eq.\ (\ref{eq-HF-eqn}) for a series of $\lambda$
values, e.g.,
$\lambda=\ldots,-2\kappa,-\kappa,0,\kappa,2\kappa,\ldots,$ (the
$\lambda=0$ solutions mean the original KS solutions) the
derivatives of orbital energies, Eq.\ (\ref{eq-PT-der-ES}), can be
found with sufficiently high accuracy by applying a
finite-difference (FD) formula of numerical differentiation:
\begin{subequations}
\label{eq-der-numES}
\begin{eqnarray}
\overset\circ\epsilon_a[\delta \hat v] &=&
^{[p,\kappa]}\overset\circ\epsilon_a[\delta \hat v]+O(\kappa^p),
\label{eq-der-numa}\\
^{[1,\kappa]}\overset\circ\epsilon_a &=&
(\epsilon_a^\kappa-\epsilon_a)/\kappa,\label{eq-der-numb}\\
^{[2,\kappa]}\overset\circ\epsilon_a &=&
(\epsilon_a^\kappa-\epsilon_a^{-\kappa})/(2\kappa),
\label{eq-der-numc}\\
^{[3,\kappa]}\overset\circ\epsilon_a &=&
(-2\epsilon_a^{-\kappa}-3\epsilon_a+6\epsilon_a^{\kappa}
-\epsilon_a^{2\kappa})/(6\kappa),
\label{eq-der-numd}\\
^{[4,\kappa]}\overset\circ\epsilon_a &=&
(8\epsilon_a^{\kappa}\!-8\epsilon_a^{-\kappa}\!
-\epsilon_a^{2\kappa}\!+\epsilon_a^{-2\kappa})/(12\kappa),
\label{eq-der-nume}
\end{eqnarray}
\end{subequations}
and so on. The derivatives of orbitals, Eq.\ (\ref{eq-PT-der-OS}),
can be found from analogous expressions for a particular
$^{[p,\kappa]}\overset\circ\phi_a$
\begin{equation}
\overset\circ\phi_a[\delta \hat v]({\bm r}) =\,
^{[p,\kappa]}\overset\circ\phi_a[\delta \hat v]({\bm
r})+O(\kappa^p)\,.\label{eq-der-numOS}
\end{equation}
Here $p$ represents the number of the HF-like equations to be
solved, and, simultaneously, the power of $\kappa$ in the error
term, see Eqs.\ (\ref{eq-der-numa}) and (\ref{eq-der-numOS}).

When performing numerical differentiation of $\phi_a^\lambda({\bm
r})$ with respect to $\lambda$, one must take care to choose
consistently phase factors of involved wave functions with various
$\lambda$. If $\phi_a^\lambda$ is a real eigenfunction of Eq.\
(\ref{eq-HF-eqn}), then $(-\phi_a^\lambda)$ is also a valid real
eigenfunction. To see if the given $\phi_a^\lambda$ is close to
$\phi_a^{0}$, let us express it as
$\phi_a^\lambda=C\,(\phi_a^{0}+\delta \phi_a^\lambda)$ (the allowed
$C$ is $+1$ or $-1$) and calculate the integral $I^\lambda=\int
\text{d}^3r\,(\phi_a^\lambda-\phi_a^{0})^2 = 2(1-C)+C\int
\text{d}^3r\,(\delta\phi_a^\lambda)^2$ ($\phi_a^{0}$ and
$\phi_a^\lambda$ were assumed normalized). We see that when
$\phi_a^\lambda$ and $\phi_a^{0}$ match ($C=+1$), the result is
$I^\lambda=\int \text{d}^3r\,(\delta\phi_\lambda)^2$, while in the
opposite case ($C=-1$), we find $I^\lambda=4-\int
\text{d}^3r\,(\delta\phi_a^\lambda)^2$. Taking a `conservative'
estimate that $0< \int \text{d}^3r\,(\delta\phi_a^\lambda)^2<2$, we
recommend to transform $\phi_a^\lambda\rightarrow -\phi_a^\lambda$
if $I^\lambda$ is found to exceed $2$. Then $\phi_a^\lambda$ and
$\phi_a^{0}$ can be partners in FD formulas.
\section{ Variational determination of model potential \protect\\
parameters}
\label{S5}
%@@@@@@@@@@@@@@@@@@@@@@@@@@
\subsection{Minimization of density-shift-finiteness indicator}
\label{S5A}
%&&&&&&&&&&&&&&&&&&&&
While the exact exchange potential $v_{\text{x}}^{\text{E}}({\bm
r})$ is defined by vanishing of the density shift in the whole
space, Eq.\ (\ref{eq-OEP}), the density shift due to some
approximate exchange potential $v_{\text{x}}^{\text{L}}({\bm r})$,
Eq.\ (\ref{eq-den-shift}), remains finite. As a convenient global
indicator of density shift finiteness (DSF), we define the
functional
\begin{equation}
\Delta_{\text{DSF}}[\delta n] = \Vert \delta n \Vert_w
\,,\label{eq-DSF}
\end{equation}
in terms of the functional norm
\begin{equation}
\Vert f \Vert_w = \biggl\{\int \text{d}^3r\,w({\bm
r})\,\biglb(f({\bm r})\bigrb)^2 \biggr\}^{1/2}\,.\label{eq-norm}
\end{equation}
The weighting function $w({\bm r}) > 0$ may be helpful in selecting
regions of special importance for the DSF indicator, e.g., $w({\bm
r})=n^p({\bm r})/\int \text{d}^3r'\,n^p({\bm r}')$  enhances
contributions in the low-density regions when $1>p>0$. We use
$w=1/N^2$ (unless stated otherwise).

Let us construct a model (m) exchange potential
$v_{\text{x}}^{\text{m}}$ as a function of some parameters. Since,
obviously,
\begin{equation}
\Delta_{\text{DSF}}\bigl[\delta\breve
n[v_{\text{x}}^{\text{m}}]\bigr]
> \Delta_{\text{DSF}}\bigl[\delta\breve
n[v_{\text{x}}^{\text{E}}]\bigr]=0\,, \label{eq-rel-DSF}
\end{equation}
the minimization of $\Delta_{\text{DSF}}\bigl[\delta\breve
n[v_{\text{x}}^{\text{m}}]\bigr]$ with respect to these parameters
leads to an improved potential, closer to $v_{\text{x}}^{\text{E}}$,
because the corresponding $\Delta_{\text{DSF}}$ becomes closer to
zero.
\subsection{Optimization of $N-1$ parameters}
\label{S5B}
%&&&&&&&&&&&&&&&&&&&&
Having some approximate exchange potential $v_{\text{x}}^{\text{L}}$
written in its canonical form, Eqs.\
(\ref{eq-identity})--(\ref{eq-vx-OS}), we are going to define a
model exchange potential $v_{\text{x}}^{\text{m}}$ by replacing the
$v_{\text{x}}^{\text{L}}$-dependent  ES
$\{D_{aa}^{\text{L}}\}_{a=1}^N$ with the model ones
$\{D_{aa}^{\text{m}}\}$ (they will play the role of variational
parameters for the mentioned minimization; at convergence, they will
become $\{D_{aa}^{\text{E}}\}$):
\begin{equation}
v_{\text{x}}^{\text{m}}({\bm r})= v_{\text{x}}^{\text{Sl}}({\bm r})+
\sum_{a=1}^{N} v_{\text{x}}^{aa}({\bm r}) \,D_{aa}^{\text{m}} +
v_{\text{x}}^{\text{OSL}}({\bm r})\,. \label{eq-vx-m}
\end{equation}
In fact, $v_{\text{x}}^{\text{m}}$ can be expressed as a corrected
$v_{\text{x}}^{\text{L}}$, namely
\begin{equation}
v_{\text{x}}^{\text{m}}({\bm r})= v_{\text{x}}^{\text{L}}({\bm r})+
\sum_{a=1}^{N} v_{\text{x}}^{aa}({\bm r}) \,d_{aa}^{\,\text{m}}
\label{eq-vx-mL}
\end{equation}
where $\{d_{aa}^{\,\text{m}}\}$ can play the role of new model
parameters, replacing previous ones by
\begin{equation}
d_{aa}^{\,\text{m}}=D_{aa}^{\text{m}}-D_{aa}^{\text{L}}\,.
\label{eq-difdaa-m}
\end{equation}
The task of finding  the model ES $\{D_{aa}^{\text{m}}\}$ that
minimize the DSF indicator $\Delta_{\text{DSF}}[\delta
n^{\text{m}}]$ for $\delta n^{\text{m}} = \delta
n\bigl[\{\delta\phi_a[v_{\text{x}}^{\text{m}} -
\hat{v}_{\text{x}}^{\text{F}}]\}\bigr]$, Eq.\ (\ref{eq-den-shift}),
is equivalent to minimization of this $\Delta_{\text{DSF}}$ with
respect to parameters $\{d_{aa}^{\,\text{m}}\}$, Eq.\
(\ref{eq-difdaa-m}), since the ES $\{D_{aa}^{\text{L}}\}$ are fixed
during the minimization.

Due to the linearity of the left-hand side of the differential Eq.\
(\ref{eq-OS-diff}) in the OS and the linearity of its right-hand
side in the potential shift, we can evaluate the total OS [yielding
$\delta n^{\text{m}}$ from  Eq.\ (\ref{eq-den-shift})] as a sum of
their constituents
\begin{subequations}
\label{eq-OS-lin}
\begin{equation}
\delta\phi_a[v_{\text{x}}^{\text{m}} -
\hat{v}_{\text{x}}^{\text{F}}]({\bm r})=
\delta\phi_a^{\text{L}}({\bm r}) + \sum_{b=1}^{N}
\delta\phi_a^{bb}({\bm r})\,d_{bb}^{\,\text{m}} \,,
\label{eq-OS-sum}
\end{equation}
where the partial OS, i.e.,
\begin{eqnarray}
\delta\phi_a^{\text{L}}({\bm r})&=&
\delta\phi_a[v_{\text{x}}^{\text{L}}-
\hat{v}_{\text{x}}^{\text{F}}]({\bm r})\,,
\label{eq-OS-Lpar} \\
\delta\phi_a^{bb}({\bm r}) &=& \delta\phi_a[v_{\text{x}}^{bb}]({\bm
r})\,, \label{eq-OS-bb}
\end{eqnarray}
are the solutions of the differential equation (\ref{eq-OS-diff}) in
which the functional argument of $\hat w_a[.]$ is the same as the
argument of $\delta\phi_a[.]$ seen in Eq.\ (\ref{eq-OS-Lpar}) or
(\ref{eq-OS-bb}). In practice, the partial OS are determined from
the solutions of the perturbed KS equations, see Sec.\ \ref{S4}.
\end{subequations}

The optimal parameters $\{d_{aa}^{\,\text{m}} =
d_{aa}^{\,\text{L}}\}$, determined by minimization, are solutions of
the set of algebraic equations $\partial \Delta_{\text{DSF}}[\delta
n]/\partial d_{aa}^{\,\text{m}}=0$, i.e.,
\begin{subequations}
\label{eq-opt-D}
\begin{equation}
\sum_{c=1}^{N} A_{ac}\,d_{cc}^{\,\text{L}} = b_a^{\text{L}}\,,\ \ \
a=1,2,\ldots,N\,, \label{eq-opt-alg}
\end{equation}
where
\begin{eqnarray}
A_{ac} &=& \int \text{d}^3r\,w({\bm r})\,\delta n^{aa}({\bm
r})\, \delta n^{cc}({\bm r}) = A_{ca}\,, \label{eq-def-A}\\
b_a^{\text{L}} &=& -\int \text{d}^3r\,w({\bm r})\,\delta n^{aa}({\bm
r})\, \delta n^{\text{L}}({\bm r})\,, \label{eq-def-B}
\end{eqnarray}
with [cf.\ Eq.\ (\ref{eq-den-shift})]
\begin{eqnarray}
\delta n^{\text{L}}({\bm r}) &=& \sum_{a=1}^{N} 2 \phi_a({\bm
r})\,\delta\phi_a^{\text{L}}({\bm r})\,, \label{eq-def-dn-L}\\
\delta n^{cc}({\bm r}) &=& \sum_{a=1}^{N} 2 \phi_a({\bm
r})\,\delta\phi_a^{cc}({\bm r})\,. \label{eq-def-dn-bb}
\end{eqnarray}
\end{subequations}
Let us note that $\sum_{b=1}^{N} A_{bc} = 0$. The following steps
lead to this result: $\sum_b \delta\phi_a^{bb} = \sum_b
\delta\phi_a[v_{\text{x}}^{bb}] = \delta\phi_a[\sum_b
v_{\text{x}}^{bb}] = \delta\phi_a[1] = 0$, due to identity
(\ref{eq-v_aa-ident}); therefore $\sum_b \delta n^{bb} = \sum_a
2\phi_a \sum_b \delta\phi_a^{bb} = 0$. With the singular coefficient
matrix, $\det(\{A_{ac}\})=0$, the solution of the system
(\ref{eq-opt-alg}) is not unique. By choosing (arbitrarily)
$d_{NN}^{\,\text{L}} = 0$ and taking $a=1,2,\ldots,(N-1)$, we obtain
a truncated system
\begin{equation}
{\bf {\sf A}}\,{\bf {\sf d}}^{\,\text{L}} = {\bf {\sf
b}}^{\,\text{L}} \label{eq-alg-N-1}
\end{equation}
having a unique solution
\begin{equation}
\bf {\sf d}^{\,\text{L}}={\bf {\sf A}}^{-1}\,{\bf {\sf
b}}^{\,\text{L}}\,. \label{eq-sol-N-1}
\end{equation}
Here ${\bf {\sf A}}$ is $(N-1)\times(N-1)$ square matrix
$\{A_{ac}\}_{a,c=1}^{N-1}$, while ${\bf {\sf b}}^{\,\text{L}}$ and
${\bf {\sf d}}^{\,\text{L}}$ are $(N-1)\times 1$ column matrices
$\{b_c^{\,\text{L}}\}_{c=1}^{N-1}$ and
$\{d_{cc}^{\,\text{L}}\}_{c=1}^{N-1}$. The possibility of arbitrary
choice for $d_{NN}^{\,\text{L}}$ reflects the fact that optimized
$v_{\text{x}}^{\text{m}}({\bm r})$ can be found only with accuracy
to an additive constant.

Finally, the approximate exchange potential
$v_{\text{x}}^{\text{L}}$ {\em optimized} with respect to $(N-1)$
parameters is
\begin{equation}
v_{\text{x}}^{\text{L},N-1}({\bm r})= v_{\text{x}}^{\text{L}}({\bm
r})+ \sum_{a=1}^{N-1} v_{\text{x}}^{aa}({\bm r})\,
d_{aa}^{\,\text{L}}\,. \label{eq-vx-opt,N-1}
\end{equation}
Because of the factual definition (\ref{eq-vx-mL}) of the model
potential, no particular (like canonical) form of
$v_{\text{x}}^{\text{L}}$ was needed to obtain the last result. At
the given step of iterations, we need to calculate only the column
${\bf {\sf b}}^{\,\text{L}}$ (as dependent on
$v_{\text{x}}^{\text{L}}$), while the matrices ${\bf {\sf A}}$ and
${\bf {\sf A}}^{-1}$ can be prepared in advance (as common to all
iterations at fixed density). From the above form of
$v_{\text{x}}^{\text{L},N-1}$ it follows that
$[v_{\text{x}}^{\text{L},N-1}({\bm r})-v_{\text{x}}^{\text{L}}({\bm
r})]$ is exponentially small in the large-$r$ region for a {\em
general} direction ${\bm r}/r$.

Once the optimized model exchange potential
$v_{\text{x}}^{\text{L},N-1}$ is found, one may be tempted to repeat
immediately such optimization procedure by starting with
$v_{\text{x}}^{\text{L},N-1}$ as a potential playing the role of
$v_{\text{x}}^{\text{L}}$. However, the corresponding new model
potential $v_{\text{x}}^{\text{m}'}$ with the  new variational
parameters  $d_{aa}^{\,\text{m}'}$ could then be represented,
according to Eq.\ (\ref{eq-vx-mL}) with $d_{NN}^{\,\text{L}}=0$, as
\begin{eqnarray}
v_{\text{x}}^{\text{m}'}({\bm r})&=&
v_{\text{x}}^{\text{L},N-1}({\bm r})+ \sum_{a=1}^{N}
v_{\text{x}}^{aa}({\bm r}) \,d_{aa}^{\,\text{m}'}
\nonumber\\
&=& v_{\text{x}}^{\text{L}}({\bm r})+ \sum_{a=1}^{N}
v_{\text{x}}^{aa}({\bm r})(d_{aa}^{\,\text{L}}
+d_{aa}^{\,\text{m}'}) \,. \label{eq-vx-mL-new}
\end{eqnarray}
Since the above potential $v_{\text{x}}^{\text{m}'}$ has exactly the
same functional form as $v_{\text{x}}^{\text{m}}$, Eq.\
(\ref{eq-vx-mL}), but with parameters $d_{aa}^{\,\text{m}}$ replaced
by $(d_{aa}^{\,\text{L}} + d_{aa}^{\,\text{m}'})$, the indicators
$\Delta_{\text{DSF}}$ corresponding to $v_{\text{x}}^{\text{m}}$ and
to $v_{\text{x}}^{\text{m}'}$ attain the same minimum, first one at
$d_{aa}^{\,\text{m}}= d_{aa}^{\,\text{L}}$, second one at
$d_{aa}^{\,\text{m}'} = d_{aa}^{\,\text{L}'} = 0$. This means that
the two optimized model potentials are identical,
$v_{\text{x}}^{\text{L}'\!,N-1}=v_{\text{x}}^{\text{L},N-1}$, so no
further improvement of $v_{\text{x}}^{\text{m}}$ can be obtained by
repeating the described minimization of $\Delta_{\text{DSF}}$.
\subsection{Optimization of $N$ parameters}
\label{S5C}
%&&&&&&&&&&&&&&&&&&&&
As an improvement to the model potential (\ref{eq-vx-m}) used in
Sec.\ \ref{S5B}, we propose to include a new parameter for
optimization --- the ``strength'' $\tilde{C}^{\text{OSm}}$ of the
model OS potential, in addition to $N$ parameters --- the ES
$\{\tilde{D}_{aa}^{\text{m}}\}_{a=1}^{N}$. Namely we define a model
exchange potential
\begin{equation}
\tilde{v}_{\text{x}}^{\text{m}}({\bm r})=
v_{\text{x}}^{\text{Sl}}({\bm r})+ \sum_{a=1}^{N}
v_{\text{x}}^{aa}({\bm r}) \,\tilde{D}_{aa}^{\text{m}} +
\tilde{C}^{\text{OSm}} v_{\text{x}}^{\text{OSL}}({\bm r})
\label{eq-vx-m-os}
\end{equation}
by replacing in Eq.\ (\ref{eq-identity}) the L-dependent objects
$\{D_{aa}^{\text{L}}\}_{a=1}^N$, $v_{\text{x}}^{\text{OSL}}({\bm
r})$, with the model ones $\{\tilde{D}_{aa}^{\text{m}}\}_{a=1}^N$,
$\tilde{C}^{\text{OSm}}v_{\text{x}}^{\text{OSL}}({\bm r})$, which,
at convergence, will become $\{D_{aa}^{\text{E}}\}$,
$v_{\text{x}}^{\text{OSE}}({\bm r})$. It  will be convenient to
change the notation
\begin{equation}
\tilde{D}_{00}^{\text{m}} \equiv \tilde{C}^{\text{OSm}}\,,\ \
v_{\text{x}}^{00}({\bm r}) \equiv v_{\text{x}}^{\text{OSL}}({\bm
r})\,,\label{eq-vx-m-00}
\end{equation}
to have the model exchange potential in a simple form
\begin{equation}
\tilde{v}_{\text{x}}^{\text{m}}({\bm r})=
v_{\text{x}}^{\text{Sl}}({\bm r})+ \sum_{a=0}^{N}
v_{\text{x}}^{aa}({\bm r}) \,\tilde{D}_{aa}^{\text{m}}\,.
\label{eq-vx-m'}
\end{equation}
which can be further rewritten as
\begin{equation}
\tilde{v}_{\text{x}}^{\text{m}}({\bm r})=
v_{\text{x}}^{\text{L}}({\bm r})+ \sum_{a=0}^{N}
v_{\text{x}}^{aa}({\bm r}) \,\tilde{d}_{aa}^{\,\text{m}}\,.
\label{eq-vx-m'-daa}
\end{equation}
where $\tilde{d}_{aa}^{\,\text{m}}$ are given analogously as in Eq.\
(\ref{eq-difdaa-m}), now valid also for $a=0$ if we set
$\tilde{D}_{00}^{\text{L}}=1$. The best model parameters are
determined by minimization of the DSF indicator
$\Delta_{\text{DSF}}[\delta n]$. The same equations as in Sec.\
\ref{S5B} are obtained, only the range of indices is extended now
from the initial $a,c=1$ to $a,c=0$. Due to singularity of the
coefficient matrix, we again choose $\tilde{d}_{NN}^{\,\text{L}} =
0$. The following (truncated) system of equations
\begin{equation}
{\bf {\sf A}}^{'}\,{\bf \tilde{\sf d}}^{'\text{L}} = {\bf {\sf
b}}^{'\text{L}} \label{eq-alg-N}
\end{equation}
is obtained, having a unique solution
\begin{equation}
{\bf \tilde{\sf d}}^{'\text{L}}=\bigl({\bf {\sf
A}}^{'}\bigr)^{-1}\,{\bf {\sf b}}^{'\text{L}}\,. \label{eq-sol-N}
\end{equation}
Here ${\bf {\sf A}}^{'}$ is $N\times N$ square matrix
$\{A_{ac}\}_{a,c=0}^{N-1}$, while ${\bf {\sf b }}^{'\text{L}}$ and
${\bf \tilde{\sf d}}^{'\text{L}}$ are $N\times 1$ column matrices
$\{b^{\,\text{L}}_c\}_{c=0}^{N-1}$ and
$\{{\tilde{d}}^{\,\text{L}}_{cc}\}_{c=0}^{N-1}$. It should be noted
that the elements $A_{0c}=A_{c0}$ and $A_{00}$ are specific to
$v_{\text{x}}^{\text{L}}$, because they depend on
$v_{\text{x}}^{00}\equiv v_{\text{x}}^{\text{OSL}}$ via $\delta
n^{00}$. Therefore, they, ${\bf {\sf b}}^{'\text{L}}$, and
$\bigl({\bf {\sf A}}^{'}\bigr)^{-1}$ need to be calculated anew at
each step of iterations. However, as shown in Appendix, costly
evaluation of $\bigl({\bf {\sf A}}^{'}\bigr)^{-1}$ at each step can
be avoided when a special method of solving Eq.\ (\ref{eq-alg-N}) is
applied, using ${\bf{\sf A}}^{-1}$  prepared in advance.

Finally, the approximate exchange potential
$v_{\text{x}}^{\text{L}}$ {\em optimized} with respect to $N$
parameters is
\begin{equation}
v_{\text{x}}^{\text{L},N}({\bm r})= v_{\text{x}}^{\text{L}}({\bm
r})+ \sum_{a=0}^{N-1} v_{\text{x}}^{aa}({\bm r})\,
{\tilde{d}}^{\,\text{L}}_{aa}\,. \label{eq-vx-opt,N}
\end{equation}
The OS potential $v_{\text{x}}^{\text{OSL}}$ of the canonical form
of $v_{\text{x}}^{\text{L}}$ is needed to obtain and write down this
result, since it is involved in  Eq.\ (\ref{eq-vx-m-00}). It should
be noted that costly evaluation of $v_{\text{x}}^{\text{OSL}}$
according to the definition (\ref{eq-vx-OS}) can be avoided, if the
identity character of Eq.\ (\ref{eq-identity}) is used. Then
\begin{equation}
v_{\text{x}}^{\text{OSL}}({\bm r})=v_{\text{x}}^{\text{L}}({\bm r})-
v_{\text{x}}^{\text{Sl}}({\bm r})-\sum_{a=1}^N
v_{\text{x}}^{aa}({\bm r}) \,D_{aa}^{\text{L}}\,. \label{eq-OSL-id}
\end{equation}
The ES $\{D_{aa}^{\text{L}}\}$ are easily evaluated as matrix
elements, Eq.\ (\ref{eq-ES-L}) with (\ref{eq-ES-def}), or with the
help of the method described in Sec.\ \ref{S4}).
\section{ Iterative algorithm leading to the exact exchange \protect\\
potential and selfconsistent density}
\label{S6}
%@@@@@@@@@@@@@@@@@@@@@@@@@@
%
\subsection{Iteration improving density: step $\ell=0$}
\label{S6A}
%&&&&&&&&&&&&&&&&&&&&&&&&
When demonstrating efficiency of our algorithm for determination of
the exact exchange potential $v_{\text{x}}^{\text{E}}$ of DFT, two
loops of iterations leading to self-consistency are employed: the
external loop for improving the density (labeled by $\{\ell\}$,
$\ell=0,1,\ldots$) and the internal loop for improving the exchange
potential (labeled by $[k]$, $k=0,1,\ldots$). At fixed $\{\ell\}$,
all iterations are performed for the given density $n^{\{\ell\}}$ of
the KS system (and, therefore, for its fixed characteristics
$\{\phi_b,\epsilon_b\},v_{\text{s}}$).

As a starting approximation to the  exact exchange potential at
$\ell =0$ we choose the Krieger-Li-Iafrate (KLI) potential
\cite{KLI92} which can be defined as having the canonical form with
the OS component equal zero
\begin{equation}
v_{\text{x}}^{\text{KLI}}[n]({\bm r}) =
v_{\text{x}}^{\text{Sl}}({\bm r}) + \sum_{a=1}^{N-1}
v_{\text{x}}^{aa}({\bm r}) \, D_{aa}^{\text{KLI}} \label{eq-vx-KLI}
\end{equation}
and the ES of the ES component satisfying
\begin{equation}
D_{aa}^{\text{KLI}} = D_{aa}\left[v_{\text{x}}^{\text{KLI}} -
\hat{v}_{\text{x}}^{\text{F}}\right] \,, \label{eq-ES-KLI}
\end{equation}
[see Eqs.\ (\ref{eq-ES-L}) and (\ref{eq-ES-def})]. Therefore
$\{D_{aa}^{\text{KLI}}\}$ can be determined from a system of
algebraic equations, $a=1,2,\ldots,N-1$,
\begin{equation}
D_{aa}^{\text{KLI}} = D_{aa}\left[v_{\text{x}}^{\text{Sl}}-
\hat{v}_{\text{x}}^{\text{F}}\right] + \sum_{b=1}^{N-1}
D_{aa}\left[v_{\text{x}}^{bb}\right]\,D_{bb}^{\text{KLI}}\,.
\label{eq-sc-KLI}
\end{equation}
Note that $D_{NN}^{\text{KLI}}\equiv \langle
\phi_N|v_{\text{x}}^{\text{KLI}}-
\hat{v}_{\text{x}}^{\text{F}}|\phi_N\rangle=0$  is satisfied.

 The KS solutions are determined next self-consistently
with this $v_{\text{x}}^{\text{KLI}}[n]$ playing the role of the
exchange contribution to the total KS potential
$v_{\text{s}}[n]=v_{\text{ext}}+v_{\text{es}}[2n]+v_{\text{x}}[n]$
(the exchange-only approximation of DFT is applied by us to
spin-compensated systems where $n=n_\uparrow=n_\downarrow$). In this
way for $\ell =0$ the starting exchange potential and KS system are
constructed with
\begin{equation}
v_{\text{x}}^{[0]}[n^{\{0\}}] =
v_{\text{x}}^{\text{KLI}}[n^{\{0\}}]\,,\ \ v_{\text{s}}^{\{0\}} =
v_{\text{ext}}+v_{\text{es}}[2n^{\{0\}}] +
v_{\text{x}}^{[0]}\,,\label{eq-vx-ini}
\end{equation}
based on the self-consistent solutions.

Before entering calculations at $(\ell=0,\ k=0)$ one should choose
values of some general parameters: $p,\kappa$ --- for numerical
differentiation in Eqs.\ (\ref{eq-der-numES}), (\ref{eq-der-numOS}),
$k_0>1$ --- the final iteration step, $\alpha_{\text{mix}}$ --- for
mixing KS potentials in  Eq.\ (\ref{eq-vs-ini_l+1}) (applied by us
value is 0.4), $\eta_{\text{DSF}}$ and $\eta_{\text{DV}}$ ---
thresholds of accuracy [see Eq.\ (\ref{eq-rel-DSF[k_0]})], $w({\bm
r})$ --- weight function used in Eq.\ (\ref{eq-norm}) and elsewhere;
$w=1/N^2$ is applied here.
\subsection{ Iteration improving exchange potential:
step~$k=0$} \label{S6B}
%&&&&&&&&&&&&&&&&&&&&&
Quantities depending only on the fixed characteristics of the actual
$\ell$th KS system are calculated:\ \ $v_{\text{x}}^{\text{Sl}}({\bm
r})$, Eq.\ (\ref{eq-vx-Slater});\ \ $v_{\text{x}}^{bb}({\bm r})$,
$\delta \phi_a^{bb}({\bm r})\equiv\delta
\phi_a[v_{\text{x}}^{bb}]({\bm r}) =
\,^{[p,\kappa]}\overset\circ\phi_a[v_{\text{x}}^{bb}]({\bm r})$,
Eqs.\ (\ref{eq-vx^aa}), (\ref{eq-der-numOS}), for $a=1,\ldots,N$,
$b=1,\ldots,N-1$ (KS-like equations are to be solved, see Sec.
\ref{S4});\ \ $\delta n^{bb}({\bm r})$, $A_{bc}$, Eqs.\
(\ref{eq-def-dn-bb}), (\ref{eq-def-A}), for $b,c=1,\ldots,N-1$;\ \
${\bf {\sf A}}^{-1}$ as reciprocal to ${\bf {\sf A}}$ --- the matrix
$\{A_{bc}\}_{b,c=1}^{N-1}$. In fact, all these calculations can be
performed at $\ell=0$ only, [see the comment below Eq.\
(\ref{eq-modES,N-1})], while at $\ell>0$ solely
$v_{\text{x}}^{\text{Sl}}({\bm r})$ and $\{v_{\text{x}}^{bb}({\bm
r})\}$ are to be calculated. Set $k=0$.
\subsection{Iteration improving exchange potential:
step~$k+1$} \label{S6C}
%&&&&&&&&&&&&&&&&&&&&&
%
\subsubsection{Initialization}
\label{S6C1}
%#################################
For given $v_{\text{x}}^{[k]}$, using Eqs.\ (\ref{eq-PT-der-OS}),
(\ref{eq-der-numOS}), (\ref{eq-PT-der-ES}) and (\ref{eq-der-numES}),
OS and ES are calculated (for $a=1,\ldots,N$)
\begin{eqnarray}
\delta \phi_a^{[k]}&\equiv&\delta
\phi_a[v_{\text{x}}^{[k]}-\hat{v}_{\text{x}}^{\text{F}}]=
\,^{[p,\kappa]}\overset\circ\phi_a[v_{\text{x}}^{[k]}
-\hat{v}_{\text{x}}^{\text{F}}]\,,\label{eq-6d-OS}\\
D_{aa}^{[k]}&\equiv&D_{aa}[v_{\text{x}}^{[k]}-\hat{v}_{\text{x}}^{\text{F}}]=
\,^{[p,\kappa]}\overset\circ\epsilon_a[v_{\text{x}}^{[k]}
-\hat{v}_{\text{x}}^{\text{F}}]\,,\label{eq-6d-ES}
\end{eqnarray}
--- the HF-like equations are to be solved, see
Sec. \ref{S4}. Such situation occurs at $k=0$. However, for $k>0$,
when the potential is known to be the sum
\begin{equation}
v_{\text{x}}^{[k]}= v_{\text{x}}^{[k-1]}+\delta
v_{\text{x}}^{[k]}\label{eq-6d-2}
\end{equation}
and the quantities $\delta \phi_a^{[k-1]}\equiv\delta
\phi_a[v_{\text{x}}^{[k-1]}-\hat{v}_{\text{x}}^{\text{F}}]$ and
$D_{aa}^{[k-1]}\equiv
D_{aa}[v_{\text{x}}^{[k-1]}-\hat{v}_{\text{x}}^{\text{F}}]$ are
available from the previous step, then
\begin{eqnarray}
\delta \phi_a^{[k]}&=&\delta \phi_a^{[k-1]}+
\,^{[p,\kappa]}\overset\circ\phi_a[\delta
v_{\text{x}}^{[k]}]\,,\label{eq-6d-3OS}\\
D_{aa}^{[k]}&=&\,D_{aa}^{[k-1]} +
\,^{[p,\kappa]}\overset\circ\epsilon_a[\delta
v_{\text{x}}^{[k]}]\,,\label{eq-6d-3ES}
\end{eqnarray}
so the less expensive KS-like equations are to be solved because the
potential $\delta v_{\text{x}}^{[k]}$ is local.

The obtained OS and ES are used in further calculations of the step.
In particular, the ES potential is determined as
\begin{equation}
v_{\text{x}}^{\text{ES}[k]}({\bm r})=\sum_{a=1}^N
v_{\text{x}}^{aa}({\bm r}) \,D_{aa}^{[k]}\,, \label{eq-vxES[k]}
\end{equation}
allowing to determine the OS potential directly, see Eq.\
(\ref{eq-OSL-id}), as
\begin{equation}
v_{\text{x}}^{\text{OS}[k]}({\bm r})=v_{\text{x}}^{[k]}({\bm r})-
v_{\text{x}}^{\text{Sl}}({\bm r})-v_{\text{x}}^{\text{ES}[k]}({\bm
r})\,. \label{eq-vxOS[k]}
\end{equation}
In this way all terms of the canonical form of
$v_{\text{x}}^{[k]}({\bm r})$ are available.

Next, a modification of $v_{\text{x}}^{[k]}({\bm r})$ is to be
calculated. If $k$ is {\em even}, the modified ES term is
determined: either via $(N-1)$-parameter optimization performed in
Sec.\ \ref{S6C2}, or via $N$-parameter one in Sec.\ \ref{S6C3}.
Otherwise (for {\em odd} $k$) the OS term is modified in Sec.\
\ref{S6C4}.
\subsubsection{Optimization of ES term: $(N-1)$-parameter version}
\label{S6C2}
%#################################
Optimization step discussed in Sec.\ \ref{S5B} is implemented. The
OS $\{\delta \phi_a^{[k]}\}$ calculated in Sec.\ \ref{S6C1} are
used. Applying Eqs.\ (\ref{eq-def-dn-L}), (\ref{eq-def-B}),
(\ref{eq-sol-N-1}) for $\text{L}=[k]$ calculate $\delta n^{[k]}$,
$\{b_a^{[k]}\}$, and finally ${\bf {\sf d}}^{[k]}$ (note that
$\{\delta \phi_a^{bb}\}$ and ${\bf {\sf A}}^{-1}$ were obtained at
the step $k=0$). The optimized potential, see Eq.\
(\ref{eq-vx-opt,N-1}),
\begin{equation}
v_{\text{x}}^{[k],N-1}({\bm r})= v_{\text{x}}^{[k]}({\bm r})+
\sum_{a=1}^{N-1} v_{\text{x}}^{aa}({\bm r})\, d_{aa}^{\,[k]}\,,
\label{eq-vx,ES,N-1}
\end{equation}
represents the $[k+1]$th potential, which is characterized by the
modifying term
\begin{equation}
\delta v_{\text{x}}^{[k+1]}({\bm r})=\sum_{a=1}^{N-1}
v_{\text{x}}^{aa}({\bm r}) d_{aa}^{\,[k]}\label{eq-modES,N-1}
\end{equation}
(here $k+1$ is odd), see Eq.\ (\ref{eq-vx[k+1]}).

It is worth noting in passing that it is sufficient to use $\delta
\phi_a^{bb}$, $\delta n^{bb}$ and ${\bf {\sf A}}^{-1}$ obtained at
the starting step $\ell=0, k=0$ (instead of calculating these
objects at the actual $\ell, k=0$), because this simplification
introduces to  ${\bf {\sf d}}^{[k]}$ an error ${\bf \delta{\sf
d}}^{[k]\{\ell\}}$ linear in a small KS potential modification
$\delta v_{\text{s}}^{\{\ell\}}({\bm r})=
v_{\text{s}}^{\{\ell\}}({\bm r})-v_{\text{s}}^{\{0\}}({\bm r})$,
resulting, therefore, in an error of the DSF indicator minimum
$\Delta_{\text{DSF}}$ that is quadratic in ${\bf \delta{\sf
d}}^{[k]\{\ell\}}$, so quadratic in $\delta
v_{\text{s}}^{\{\ell\}}$. We observe no appreciable effect of this
simplification on the convergence rate.

Go to  Sec.\ \ref{S6C5} --- the closing subsection.
\subsubsection{Optimization of ES term: $N$-parameter version}
\label{S6C3}
%#################################
Alternatively to previous Sec.\ \ref{S6C2}, the optimization step
discussed in Sec.\ \ref{S5C} is implemented here. The OS $\{\delta
\phi_a^{[k]}\}$ and OS potential $v_{\text{x}}^{\text{OS}[k]}$,
calculated in Sec.\ \ref{S6C1}, are used. For $\text{L}=[k]$
calculate  $v_{\text{x}}^{00}$, Eq. (\ref{eq-vx-m-00}); $\{\delta
\phi_a^{00}\equiv\delta \phi_a[v_{\text{x}}^{00}]=
\,^{[p,\kappa]}\overset\circ\phi_a[v_{\text{x}}^{00}]\}_{a=1}^N$,
Eqs.\ (\ref{eq-PT-der-OS}), (\ref{eq-der-numOS}); $\delta n^{00}$,
Eq.\ (\ref{eq-def-dn-bb}); $A_{0c}=A_{c0}$, $A_{00}$, Eq.\
(\ref{eq-def-A}); $b_0^{[k]}$, Eq.\ (\ref{eq-def-B}).

Solve the system of equations
\begin{equation}
{\bf {\sf A}}^{'}\,{\bf \tilde{\sf d}}^{'[k]} = {\bf {\sf b}}^{'[k]}
\label{eq-6e-3}
\end{equation}
for unknown vector ${\bf \tilde{\sf d}}^{'[k]}$ [see definitions of
${\bf {\sf A}}^{'}$, ${\bf {\sf b}}^{'[k]}$, ${\bf \tilde{\sf
d}}^{'[k]}$ below Eq.\  (\ref{eq-sol-N})]. Inexpensive method of
solving this system is shown in Appendix, where the matrix ${\bf
{\sf A}}^{-1}$ (calculated once at the $k=0$ step) is used.

The optimized potential, see Eq.\ (\ref{eq-vx-opt,N}),
\begin{equation}
v_{\text{x}}^{[k],N}({\bm r})= v_{\text{x}}^{[k]}({\bm r})+
\sum_{a=0}^{N-1} v_{\text{x}}^{aa}({\bm r})\, {\tilde
d}_{aa}^{\,[k]}\,, \label{eq-vx,ES,N}
\end{equation}
represents the $[k+1]$th potential, which is characterized by the
modifying term
\begin{equation}
\delta v_{\text{x}}^{[k+1]}({\bm r})=\sum_{a=0}^{N-1}
v_{\text{x}}^{aa}({\bm r}) {\tilde d}_{aa}^{\,[k]}
\label{eq-modES,N}
\end{equation}
(here $k+1$ is odd). The simplification mentioned below Eq.\
(\ref{eq-modES,N-1}) concerning $\delta \phi_a^{bb}$, $\delta
n^{bb}$ and ${\bf {\sf A}}^{-1}$ is applicable here too.

Warning: at the starting step $(\ell=0,\ k=0)$ the $N$-parameter
optimization is impossible because
$v_{\text{x}}^{00}=v_{\text{x}}^{\text{OS\,KLI}}=0$. This is a
specific property of the chosen initial exchange potential
$v_{\text{x}}^{[0]}=v_{\text{x}}^{\text{KLI}}$, see Eq.\
(\ref{eq-vx-KLI}). The $(N-1)$-parameter optimization should be
applied instead.

Go to  Sec.\ \ref{S6C5} --- the closing subsection.
\subsubsection{Modification of OS term}
\label{S6C4}
%#################################
The recurrence relation, Eq.\ (\ref{eq-vx-CH-recur}), proposed
earlier in Sec.\ \ref{S3D}, is implemented. We transform the OS
$\delta \phi_a^{[k]}$ [obtained in Sec.\ \ref{S6C1}] into the {\em
modified} OS according to Eq.\ (\ref{eq-OSperp}), $\forall {\bm r}$,
for $a=1,\ldots,N$:
\begin{equation}
\delta\phi_a^{[k]\perp}({\bm r}) = \delta\phi_a^{[k]}({\bm r}) -
\phi_{a}({\bm r})\,\delta n^{[k]}({\bm r})/\bigl(2 n^{\{\ell\}}({\bm
r})\bigr)\,, \label{eq-OS[k]perp}
\end{equation}
to obtain the new OS potential
\begin{equation}
v_{\text{x}}^{\text{OS}[k+1]}({\bm r}) =
v_{\text{x}}^{\text{OSb}}[\{\delta\phi_a^{[k]\perp}\}\,,0\,]({\bm
r}) \label{eq-vx-OS[k+1]}
\end{equation}
[Eq.\ (\ref{eq-vx-OS-fb})], for the use in the recurrence relation
\begin{equation}
v_{\text{x}}^{[k+1]}({\bm r}) = v_{\text{x}}^{\text{Sl}}({\bm r}) +
v_{\text{x}}^{\text{ES}}[\{D_{aa}^{[k]}\}]({\bm r}) +
v_{\text{x}}^{\text{OS}[k+1]}({\bm r})\,. \label{eq-6f-recur}
\end{equation}
The modifying term is therefore
\begin{equation}
\delta v_{\text{x}}^{[k+1]}({\bm r}) =
v_{\text{x}}^{\text{OS}[k+1]}({\bm r}) -
v_{\text{x}}^{\text{OS}[k]}({\bm r}) \label{eq-modOS}
\end{equation}
(here $k+1$ is even). Note that the OS potential
$v_{\text{x}}^{\text{OS}[k]}$ was calculated in Sec.\ \ref{S6C1}.
\subsubsection{Closing}
\label{S6C5}
%#################################
The new exchange potential is obtained as
\begin{equation}
v_{\text{x}}^{[k+1]}({\bm r}) = v_{\text{x}}^{[k]}({\bm r}) +\delta
v_{\text{x}}^{[k+1]}({\bm r})\,.\label{eq-vx[k+1]}
\end{equation}

The steps for optimization of the ES term and modification of the OS
term have implemented fully our new ideas how to update the
approximate exchange potential $v_{\text{x}}^{[k]}({\bm r})$.
Therefore, these steps should be repeated to gain further
improvement. Due to the modification of the OS potential,
optimization of the model ES parameters can be effective in the next
$k$ step.

Perform the replacement $k:=k+1$ to update the step number. If $k
\ne k_0$,  go to Sec.\ \ref{S6C} to perform a consecutive $k$ step,
otherwise continue as below.
\subsection{Iteration improving density: step $\ell+1$, termination}
\label{S6D}
%&&&&&&&&&&&&&&&&&&&&&&&&
When the exchange potential iterations are terminated at $k= k_0$
(quite small values can be taken, like $k_0=2$ or $k_0=3$) the
accuracy of $v_{\text{x}}^{[k]}$ may be still low, but it is already
sufficient to improve the density in the next, $(\ell+1)$th step.
This accuracy can be measured by (smallness of)
\begin{equation}
\Delta_{\text{DSF}}^{[k]} \equiv \Delta_{\text{DSF}}[\delta n^{[k]}]
\label{eq-DSF[k0]}
\end{equation}
[see Eq.\ (\ref{eq-DSF})], where $\delta n^{[k]}$ is calculated from
$\{\delta \phi_a^{[k]}\}$ (obtained in Sec.\ \ref{S6C1}) by applying
Eq.\ (\ref{eq-def-dn-L}) for $\text{L}=[k]$.

In order to fix the arbitrary additive constant of the obtained
$v_{\text{x}}^{[k_0]}$ in agreement with Eq.\ (\ref{eq-ES-repl}),
calculate [see Eqs.\ (\ref{eq-ES-L}) and (\ref{eq-ES-def})]
\begin{equation}
D_{NN}^{[k_0]}= \langle \phi_N|v_{\text{x}}^{[k_0]}-
\hat{v}_{\text{x}}^{\text{F}}|\phi_N\rangle\,, \label{eq-D_NN[k0]}
\end{equation}
and, next, noting that $\sum_{a=1}^N v_{\text{x}}^{aa}({\bm r})=1$,
perform the replacement
\begin{equation}
v_{\text{x}}^{[k_0]}({\bm r}):=v_{\text{x}}^{[k_0]}({\bm r}) -
D_{NN}^{[k_0]}\,. \label{eq-vx[k0]shift}
\end{equation}

For the $(\ell+1)$th step, the potential
$v_{\text{x}}^{[k_0]}[n^{\{\ell\}}]({\bm r})$ is used to construct
the new KS potential as a mixture of potentials (regulated by a
parameter $0 \le \alpha_{\text{mix}} \le 1$)
\begin{eqnarray}
v_{\text{s}}^{\{\ell+1\}} &=& \alpha_{\text{mix}}\left(
v_{\text{ext}}+v_{\text{es}}[2n^{\{\ell\}}] +
v_{\text{x}}^{[k_0]}[n^{\{\ell\}}]\right) + \nonumber\\
&&(1-\alpha_{\text{mix}})\,v_{\text{s}}^{\{\ell\}}\,.
\label{eq-vs-ini_l+1}
\end{eqnarray}
The KS solutions with this KS potential are obtained. They define
the density $n^{\{\ell+1\}}({\bm r})$.

At this point one should check, using some criteria, if the density
$n^{\{\ell+1\}}$ approximates the selfconsistent density within some
presumed accuracy limits. In our example for this check (discussed
later on), the indicator of density values (DV) discrepancy
$\Delta_{\text{DV}}[n_1,n_2]=\Vert n_1-n_2 \Vert_w $ is introduced
and
$\Delta_{\text{DV}}^{\{\ell+1\}\{\ell\}}=\Delta_{\text{DV}}[n^{\{\ell+1\}},
n^{\{\ell\}}]$ is calculated. If
\begin{equation}
(\Delta_{\text{DSF}}^{[k_0]} < \eta_{\text{DSF}})\wedge
(\Delta_{\text{DV}}^{\{\ell+1\}\{\ell\}} < \eta_{\text{DV}})
\label{eq-rel-DSF[k_0]}
\end{equation}
is {\em true}, iterations over $\ell$ are to be {\em terminated}.
Here the parameters $0< \eta_{\text{DSF}}\ll 1$ and $0<
\eta_{\text{DV}}\ll 1$ represent the chosen thresholds of accuracy.
The determined exact exchange potential is
$v_{\text{x}}^{\text{E}}({\bm r}) \approx v_{\text{x}}^{[k_0]}({\bm
r})$ for all ${\bm r}$ [the correct large-$r$ behavior is guaranteed
due to the replacement (\ref{eq-vx[k0]shift})].

But if the  expression (\ref{eq-rel-DSF[k_0]}) is {\em false},
iterations over $\ell$ are to be {\em continued}. Define the initial
exchange potential as
\begin{equation}
v_{\text{x}}^{[0]}[n^{\{\ell+1\}}]({\bm r}) =
v_{\text{x}}^{[k_0]}[n^{\{\ell\}}]({\bm r})\,,\label{eq-vx-ini_l+1}
\end{equation}
perform the replacement $\ell:=\ell+1$, and go to Sec.\ \ref{S6B}
--- the exchange potential iteration step $k=0$.

\section{Results of tests}
\label{S7}

\subsection{Numerical method}
\label{S7-num-meth}

The algorithm described in the previous section is successfully
applied  to several spin-compensated closed-(sub)shell atoms, from Be up to Kr.
Similarly as in our previous work \cite{CH07}, the calculations are performed
using the highly accurate pseudospectral (PS) method in which  KS and HF orbitals
$\phi_a ({\bm r}) = R_{nl}(r)Y_{lm}(\hat{\bm r})$
(where $a=(nlm)$ is specified with the main $n$, angular momentum $l$
and magnetic $m$ quantum numbers)
are represented with the values of the
radial functions $\chi_{nl}(r)=r\,R_{nl}(r)$ at $M+2$ discrete points (nodes)
%$r_i$ ($i=0,\ldots,M$).
$r_i=r(x_i)$ which correspond to an increasing sequence $x_0=-1, x_1, \ldots, x_M, x_{M+1}=1$
via the scaling $r(x)=L(1+x)/(1-x)$ (the parameter $L=0.3$ is used).
Simultaneously, the KS- and HF-like  equations
are transformed into eigenvalue algebraic equations
for the orbital energies $\epsilon_{nl}$ and the corresponding
eigenvectors $[u_i^{nl}] \equiv [g(r_i)\chi_{nl}(r_i)]_{i=1}^M$
which  are  solved using standard techniques;
note that $\chi_{nl}(r_i)$=0 for $r_0=0$ and  $r_{M+1}=\infty$.
The form of orbital-independent function $g(r)$ and other
details on the application of the PS method
to the KS equation are given in Refs. \onlinecite{WCL94,RC01}.
With the presently chosen scaling $r(x)$,
% mapping $x \in [-1,1] \mapsto r \in [0,\infty]$,
it is also possible to fully implement the  PS representation
in numerical  solution of the HF(-like) equation for atoms, i.e.,
to use --- like for the KS equation ---
solely the  radial function values $\{\chi_{nl}(r_i)\}_{i=1}^M$  at the nodes.
This novel technique, to be reported elsewhere \cite{MC11},
differs from Friesner's approach  \cite{Fr85,Fr86}
where the HF equation is solved by combined use of
orbital basis and the
PS representation of applied functions.
In the present calculations applying the PS method,
 a very moderate number of $M=120$ nodes is used  and it is found to be sufficient for
the accuracy $10^{-10}\; \text{Hartree}$ of the occupied orbital energies
obtained for given one-electron Hamiltonian
(it is true for  both the KS(-like) and HF-like equations)
while  the self-consistent orbital energy values are
two or more orders less accurate due to inaccuracies of $v_{\text{x}}$
of the FD origin (even at convergence) as it is discussed later on.

In the PS method
the exchange potential $v_{\text{x}}(r)$ needs to be known only
at the nodes $r=r_i$ ($i=1,\ldots,M$).
Thus, it is also true for
the  constituent terms of $v_{\text{x}}(r)$  in the canonical representation and
to its increments determined in the iteration algorithm.
The ES term  $v_{\text{x}}^{\text{ES}}(r_i)$ is readily calculated
with  $\{\chi_{nl}(r_i)\}$, $nl\in \text{occ}$, while the Slater
potential  $v_{\text{x}}^{\text{Sl}}(r_i)$
is obtained with  all values $\{\chi_{nl}(r_j)\}$
 when expressing
the action of the Fock exchange operator on the orbitals in the PS representation \cite{MC11}.
Also the radial OS functions $\delta\chi_{nl}(r)$ and
their derivatives $d[\delta\chi_{nl}(r)]/dr$, used to modify the OS term of
the iterated $v_{\text{x}}(r)$  (Sec. \ref{S6C4}),
need to be determined only at $r=r_i$.
This is done with the (modified) FD formula
(\ref{eq-der-numOS}) using [instead of $\phi_{a}^{\lambda}(r_i)$]
the values
$\chi_{nl}^{\lambda}(r_i)$ which are found directly by diagonalizing
the matrix of the HF-like hamiltonian (\ref{eq-HF-ham}).
In calculation of the Slater, ES and OS terms of $v_{\text{x}}(r)$,
 the KS functions $\chi_{nl}(r)$ --- which are not accurately represented
 for large $r$ in the PS method  \cite{CH07} ---
are approximated   for $r >r_{01}$
with the asymptotic form
$\chi_{nl}^{\text{(as)}}(r)=C_{nl} r^{1/\beta_{nl}}e^{-\beta_{nl} r}$ where
$\beta_{nl}=\sqrt{-2\epsilon_{nl}}$;
the  parameters $C_{nl}$ are found by matching
$\chi_{nl}(r)$ and $\chi_{nl}^{\text{(as)}}(r)$ at
the radius $r=r_{01}$,
chosen  by our code individually for each orbital $(nl)$
(e.g., for Ne,  $r_{01}=2.85,\,11.76,\,16.89\,\text{a.u.}$ for $(nl)=1s,\, 2s, \, 2p$).
The effect of similar inaccuracies of the OS determined with the FD formula
at large $r$ can be avoided by setting
the exponentially  decaying  OS term $v_{\text{x}}^{\text{OS}}({\bm r})$
 to 0 for $r$ larger than some cut-off radius $r_{02}$
 (e.g.,  $9,\,12,\,14\,\text{a.u.}$  for Ne, Ar and Zn),
without affecting  the high accuracy of the calculated orbitals
and their energies \cite{CH07}.

The PS nodes $r_i$ are also used in
(a variant of) the highly accurate Gauss-Legendre quadrature
applied
to calculate integrals representing various quantities applied
in our method,
including the DSF and DV indicators or  the  elements of
the matrix $\bf \sf A$  and the vector
${\bf \sf b}^{\,\text{L}}$
used in the ES optimization.

\subsection{Efficiency of present algorithm for exact exchange potential}
\label{S7-test-vx}

\begin{figure}[ht!]
\includegraphics*[width=9cm]{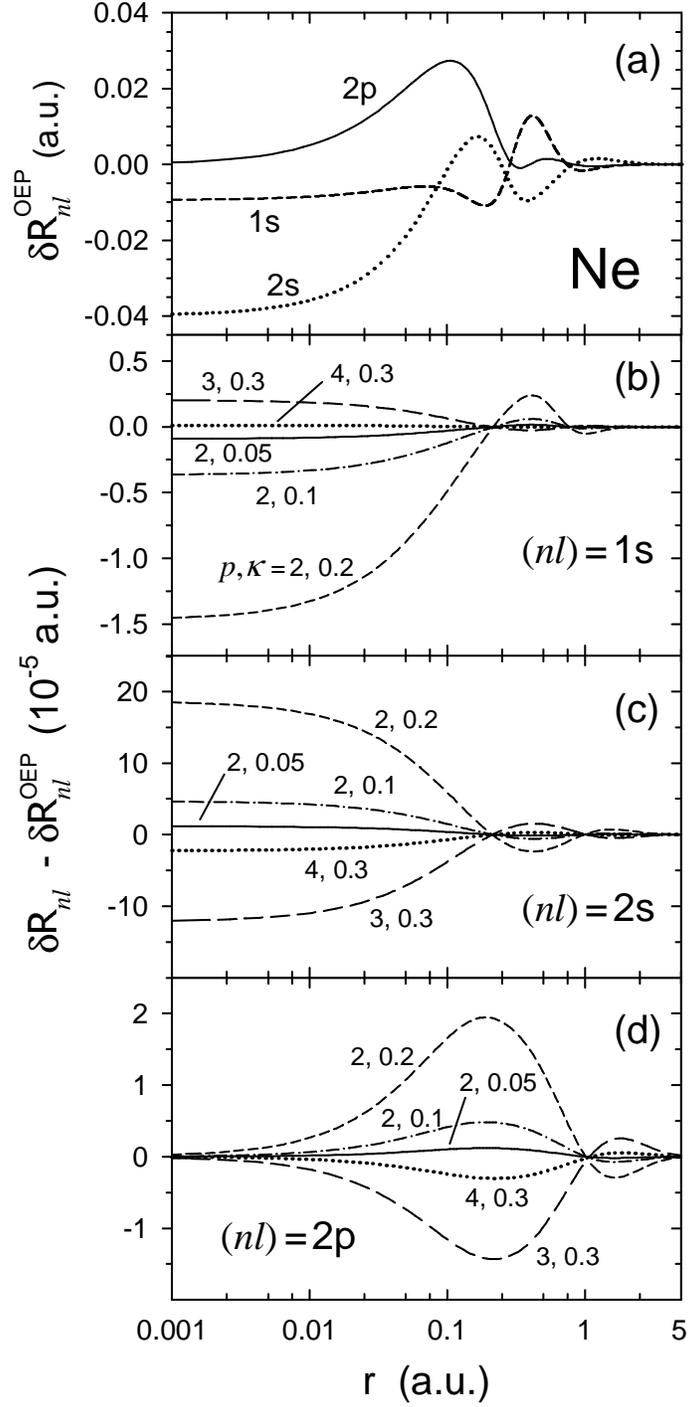}
\caption{ Approximate radial OS  $\delta R_{nl}=
\delta\chi_{nl}/r=\,^{[p,\kappa]}\overset\circ\chi_{nl}/r$
calculated for  the Ne atom with the FD formulas for various $p,\kappa$
and compared with
the exact OEP OS $\delta R_{nl}^{\text{OEP}}$, all  obtained
for the the exchange-only OEP density $n^{\text{OEP}}$
and the corresponding
exact exchange potential $v_{\text{x}}^{\text{E}}[n^{\text{OEP}}]$
determined  with
the non-iterative method of Ref.\ \onlinecite{CH07}.
 }
\label{fig-Ne_dRad-approx}
\end{figure}

To test one of the key elements of the present algorithm, the  approximate ES
$\delta\epsilon_{nl}\equiv D_{nl,nl} =\, ^{[p,\kappa]}\overset\circ\epsilon_{nl}$ and
radial OS $\delta R_{nl}=\,^{[p,\kappa]}\overset\circ R_{nl}$
are calculated for the
exchange-only OEP density $n^{\text{OEP}}$ and the corresponding exact exchange potential
$v_{\text{x}}^{\text{E}}[n^{\text{OEP}}]$ (as well as the respective orbitals and
their energies) by using the FD formulas
(\ref{eq-der-numES}) and (\ref{eq-der-numOS}) with different $p$ and
$\kappa$.
The obtained results are compared --- in Table
\ref{tab-ES-Ne} and Fig. \ref{fig-Ne_dRad-approx} ---
with the exact ES and OS, i.e.,
$\delta\epsilon_{nl}^{\text{E}} \equiv D_{nl,nl}^{\text{OEP}}$ and
$ \delta R_{nl}^{\text{OEP}}$
determined  with our method of Ref. \onlinecite{CH07}.
For given order $p$ of
the FD formula (\ref{eq-der-numES}), the approximate ES
$\delta\epsilon_{nl}=^{[p,\kappa]}
\overset\circ\epsilon_{nl}[v_{\text{x}}^{\text{E}}-\hat{v}_{\text{x}}^{\text{F}}]$
and  OS
$\delta R_{nl}=\, ^{[p,\kappa]}
\overset\circ R_{nl}[v_{\text{x}}^{\text{E}}-\hat{v}_{\text{x}}^{\text{F}}]$
become closer
to $\delta\epsilon_{nl}^{\text{E}}$ and $\delta R_{nl}^{\text{E}}$
when $\kappa$ becomes smaller.
For $p=1$ and $p=2$, the discrepancies
$\delta\epsilon_{nl}-\delta\epsilon_{nl}^{\text{E}}$
well obey the  predicted power law $\kappa^p$,
but they  decrease so quickly with increasing $p$
for each considered $\kappa=0.05,\;0.1,\;0.2$ (Table \ref{tab-ES-Ne})
that for $p=3$, $p=4$ they approach the accuracy limit,
and, in consequence, their $\kappa^p$ dependence is not clearly
seen for these $p$.
The accuracy of the calculated  ES $\delta\epsilon_{nl}$
(and their accuracy limit)
is a combined effect of the $\kappa^p$ {\em non-numerical} error of
the corresponding FD formula (\ref{eq-der-numES}) and
{\em numerical} inaccuracies  of
$\epsilon_{nl}^{\lambda}$
leading  to  an additional error in the ES,
proportional to $1/\kappa$ (which is a factor present in
FD formula (\ref{eq-der-numES})).
Similar conclusions hold for the accuracy of the OS calculated
with the  FD formula (\ref{eq-der-numOS}).
Hereafter, the OS and ES (as well as their increments) used in
determination of $v_{\text{x}}^{[k]}$ are normally calculated with
the FD formulas (\ref{eq-der-numES}), (\ref{eq-der-numOS}) for
$p=2$, $\kappa=0.05$
(invoking only two terms $\epsilon_{nl}^{\lambda}$
or $R_{nl}^{\lambda}$,  with $\lambda=-\kappa,\,\kappa$),
unless  the dependence of the result accuracy
on $p$ and $\kappa$ is studied, as in Tables \ref{tab-ES-Ne},
\ref{tab-eorb-Zn} and Fig. \ref{fig-Ne_dRad-approx}.
Note that the FD formulas of higher order $p$ (3 or 4)
%--- more accurate (in exact arithmetics) but
--- though requiring calculation of more terms
$\epsilon_{nl}^{\lambda}$ and $R_{nl}^{\lambda}$ ---
should be applied when inaccuracies of these terms become larger
since this adverse effect can be compensated with use of larger $\kappa$
without losing the accuracy of the resultant ES and OS.

\begin{figure}
\includegraphics*[width=9cm]{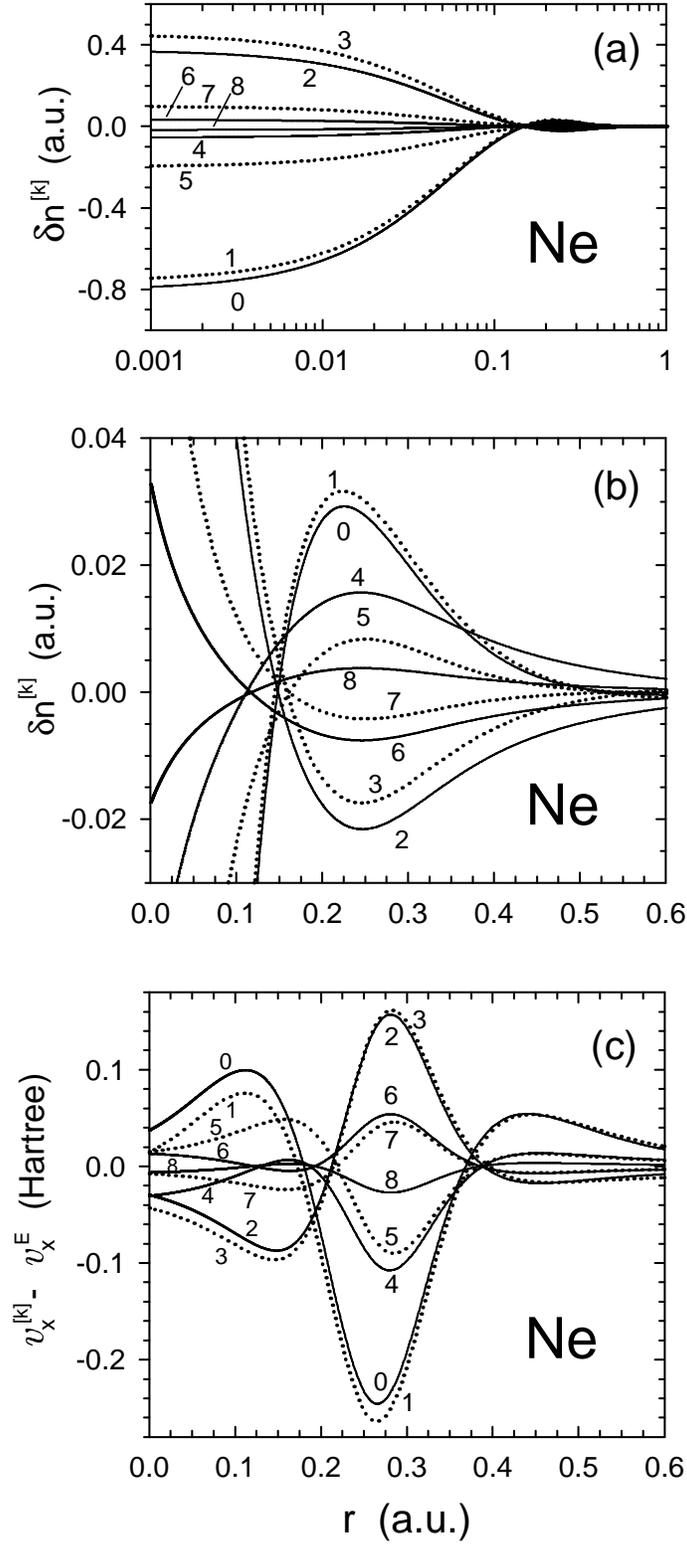}
\caption{ (a,b) Density shift $\delta n^{[k]}$ for the approximate exchange potential
$v_{\text{x}}^{[k]}$ ,
(c) difference between
$v_{\text{x}}^{[k]}$ and  the exact exchange potential
$v_{\text{x}}^{\text{E}}$,  all obtained for the the exchange-only OEP density
$n^{\text{OEP}}$ in the Ne atom.
The plots are shown for $k=0,2,4,6,8$ (solid lines)
and $k=1,3,5,7$ (dotted lines) and each of them is labeled with the corresponding $k$ value.
For even  $k$, the respective $v_{\text{x}}^{[k]}$ is obtained by modifying the OS term
in $v_{\text{x}}^{[k-1]}$
while for odd $k$  it is found by $(N-1)$-parameter optimization of the ES of
$v_{\text{x}}^{[k-1]}$.
For better comparison with $v_{\text{x}}^{\text{E}}$
an additional constant shift is applied to $v_{\text{x}}^{[k]}$
at each iteration step $k$ to obtain $D_{NN}^{[k]}=0$,
the same as $D_{NN}^{\text{E}}=0$.
}
\label{fig-Ne_dn_dvx}
\end{figure}

\begin{figure}
\includegraphics*[width=9cm]{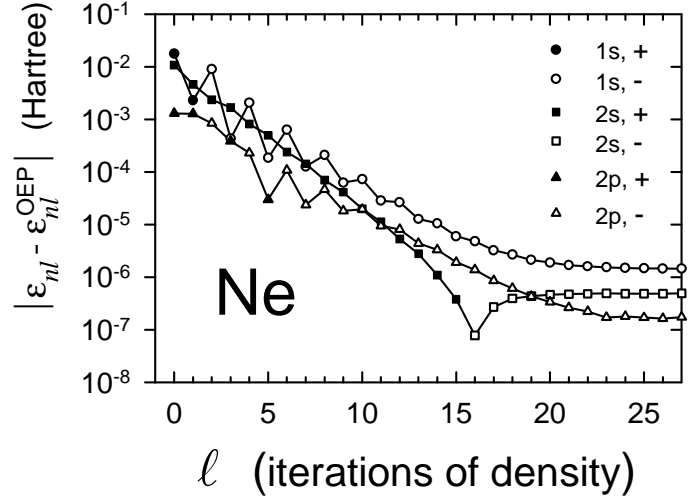}
\caption{The absolute differences between the approximate
orbital energies  $\epsilon_{nl}$
obtained for the Ne atom with
the KS potential  $v_{\text{s}}^{\{\ell\}}$, Eq. (\ref{eq-vs-ini_l+1}),
at the $\ell$-th density iteration step,
and  the respective energies $\epsilon_{nl}^{\text{OEP}}$
(equal to $-30.8200039329,\,  -1.7181258001,\, -0.8507101622\, \text{Hartree}$
for $(nl)=1s,2s,2p$)
found with the exact exchange potential  for $n^{\text{OEP}}$.
The exchange potential $v_{\text{x}}^{[k_0]}$
entering  $v_{\text{s}}^{\{\ell\}}$ is found (for $n^{\{\ell-1\}}$)
with  $k_0=2$ steps of   iteration  which employs
the $(N-1)$-parameter optimization of the  ES.
 The results for $nl=1s$ (circles), $2s$ (squares) and $2p$ (triangles)
are marked with solid symbols for positive values
 of $\epsilon_{nl}-\epsilon_{nl}^{\text{OEP}}$ and with open symbols
for negative ones. }
\label{fig-Ne_eorb}
\end{figure}

The efficiency of the  algorithm for self-consistent determination
of $n$ and $v_{\text{x}}$
has been tested for closed-(sub)shell atoms from Be to Kr
and the selected results for the Ne, Ar and Zn atoms  are
presented in  Figs. \ref{fig-Ne_dn_dvx}--\ref{fig-Ne_eorb}
and Tables \ref{tab-dv-Zn}, \ref{tab-eorb-Zn}.
The quality of the approximate exchange potential $v_{\text{x}}^{[k]}$
obtained for a given electron density ($n^{\text{OEP}}$ is used in this test)
is first assessed for the Ne atom
by direct comparison of $v_{\text{x}}^{[k]}$
with the exact  exchange potential $v_{\text{x}}^{\text{E}}[n^{\text{OEP}}]$
(its plot is given in Fig.\ 1 of Ref.\ \onlinecite{KP03}(a)).
The iterated potential $v_{\text{x}}^{[k]}$
converges quickly to $v_{\text{x}}^{\text{E}}$
(Fig. \ref{fig-Ne_dn_dvx}(c)),
so that  (the maximum magnitude of) the  discrepancy
$\Delta v_{\text{x}}^{[k]}\equiv v_{\text{x}}^{[k]}-v_{\text{x}}^{\text{E}}$
 is reduced
10, $10^2$, $10^3$ and $10^4$ times after $k=9,\,16,\,22,\,29$
iteration steps, respectively,
in comparison to  $\Delta v_{\text{x}}^{[0]}=
v_{\text{x}}^{\text{KLI}}-v_{\text{x}}^{\text{E}}$
(equal to $-0.25\,\text{Hartree}$ for $r=0.26\,\text{a.u.}$).
Largest discrepancies $|\Delta v_{\text{x}}^{[k]}|$
are found  in the interval $0.2\,\text{a.u.} <r< 0.35\,\text{a.u.}$
corresponding to the $KL$ intershell region
where the characteristic bump in the $r$-dependence of
$v_{\text{x}}^{\text{E}}$ is located
in the Ne atom.
In this region the iterated potentials $v_{\text{x}}^{[k]}$
change most significantly at even iteration steps $k$
when $v_{\text{x}}^{[k]}$ is obtained
by modifying the OS term of $v_{\text{x}}^{[k-1]}$ --- this leads
the sign change of $\Delta v_{\text{x}}^{[k]}$ in comparison to
$\Delta v_{\text{x}}^{[k-1]}$.
The $(N-1)$-parameter ES optimization of $v_{\text{x}}^{[k-1]}$ ---
applied in this test for Ne to calculate $v_{\text{x}}^{[k]}$ at odd steps $k$ ---
has much smaller effect in the intershell region.
However,
it leads to much larger changes of the exchange potential $v_{\text{x}}^{[k]}$
for $0\leq r \leq 0.2\,\text{a.u.}$ (the $K$ shell region)
which may be  comparable to those induced by the OS modification at even iteration steps.
The figure \ref{fig-Ne_dn_dvx}(a),(b) demonstrates that
diminishing of extremal $|\delta n^{[k]}|$
is in accord with diminishing
of extremal $|\Delta v_{\text{x}}^{[k]}|$, Fig.\ \ref{fig-Ne_dn_dvx}(c).

In practical calculations, when the exact potential
$v_{\text{x}}^{\text{E}}$ is not known beforehand and
thus cannot be used as a reference,
the assessment of the $v_{\text{x}}^{[k]}$ quality
(suggested by the comparison of Fig.\ \ref{fig-Ne_dn_dvx}(c)
with Fig.\ \ref{fig-Ne_dn_dvx}(a),(b))
is done with aid of the DSF indicator
$\Delta_{\text{DSF}}^{[k]}=\Vert \delta n^{[k]} \Vert$,
Eq. (\ref{eq-DSF[k0]}),
which is a suitable measure of how well
the OEP equation (\ref{eq-OEP})  is satisfied
by $v_{\text{x}}^{[k]}$ approximating $v_{\text{x}}^{\text{E}}$.
In tests for Ne, Ar, and Zn (Fig.\ \ref{fig-Ne_dsf_noep}),
the  indicator $\Delta_{\text{DSF}}^{[k]}$ calculated for a fixed density
($n^{\text{OEP}}$ is used) decreases rapidly during
iteration,
roughly as $c k^{-\gamma}$
($c$, $\gamma$ --- constants),
until a minimum level $\Delta_{\text{DSF}}^{\text{min}}$ is reached
after some number $k_1$ of iteration steps.
This corresponds to obtaining convergence of  $v_{\text{x}}^{[k]}$ for
$k=k_1$  since  $\Delta_{\text{DSF}}^{[k]}$
changes  in oscillatory manner for $k>k_1$,
with amplitude of the order of  $\Delta_{\text{DSF}}^{\text{min}}$.
The decrease of  $\Delta_{\text{DSF}}^{[k]}$ is faster
for the $N$-parameter ES optimization than for its $(N-1)$-parameter variant
though the very similar $\Delta_{\text{DSF}}^{\text{min}}$ level
is achieved in both cases.
Its value  depends on
the accuracy of the ES and OS obtained with the  FD formulas and
it is around  $10^{-7}$ for presently used $p=2$, $\kappa=0.05 $ but it can be reduced
below $10^{-8}$  for suitably chosen $p$, $\kappa$
(e.g., $p=2$, $\kappa=0.02$ or $p=4$, $\kappa=0.1$).
The number of steps  required to obtain
$\Delta_{\text{DSF}}^{[k]} \approx  10^{-7} $,
corresponding to convergent $v_{\text{x}}^{[k]}$ in the present test, is
$k_1=30,\, 17,\, 21$ with $(N-1)$-parameter optimization and
$k_1=13,\, 13,\, 11$ with $N$-parameter optimization
for the Ne, Ar, and Zn atoms, respectively.
However, note that the found difference  in the speed of
convergence of $v_{\text{x}}^{[k]}$
%of $\Delta_{\text{DSF}}^{[k]}$,
%and, consequently, of underlying  $v_{\text{x}}^{[k]}$
--- especially evident, in our test, for Ne and Zn --- occurs only when we consider
$v_{\text{x}}^{[k]}$ calculated for a given density.
This advantage of $N$-parameter ES optimization is not retained  in the
practical calculations when both the electron density and
the exchange  potential
are found iteratively; see later on.

Note also that for all {\em even} $k$
the values  $\Delta_{\text{DSF}}^{[k]}$ are smaller than
$\Delta_{\text{DSF}}^{[k-1]}$ (see Fig. \ref{fig-Ne_dsf_noep})
since the corresponding $v_{\text{x}}^{[k]}$ are obtained by minimizing
$\Delta_{\text{DSF}}$ of $v_{\text{x}}^{[k-1]}$
 by treating the ES (and $D_{00}$) of this potential
as $N-1$ (or $N$) variational parameters.
For {\em odd} $k$, we find that $\Delta_{\text{DSF}}^{[k]}$
are often larger than $\Delta_{\text{DSF}}^{[k-1]}$, despite the  mentioned overall
rapid decrease of $\Delta_{\text{DSF}}^{[k]}$.

If the ES  optimization is not used,
the corresponding DSF indicator
(marked with triangles in Fig.\ \ref{fig-Ne_dsf_noep})
 does not decrease with $k$
(or decrease very slowly, e.g., for Zn)
so that the iteration of $v_{\text{x}}^{[k]}$
does not converge.
In particular,  it is found
that, with increasing $k$, the potential $v_{\text{x}}^{[k]}$
oscillates around $v_{\text{x}}^{\text{E}}$ in the $KL$ intershell region,
changing the sign of $\Delta v_{\text{x}}^{[k]}$
after every the OS modification step, but
$|\Delta v_{\text{x}}^{[k]}|$  does not decrease during iteration.
This adverse behavior cannot be remedied with (often used in similar cases)
mixing of the current $v_{\text{x}}$ with the one obtained in the previous step:
thus the inclusion of the ES optimization step
becomes essential to obtain convergence of the iterated exchange potential.

\subsection{Efficiency of present algorithm for self-consistent
determination of density}
\label{S7-test-vx-dens}

\begin{figure}
\includegraphics*[width=9cm]{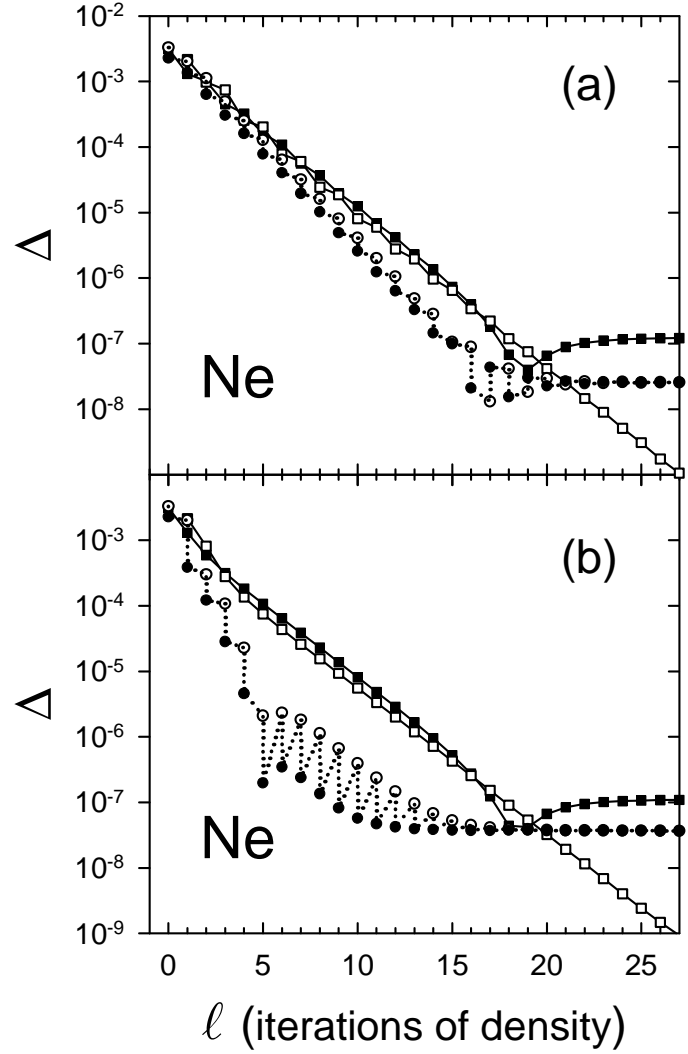}
\caption{ DSF indicators
$\Delta_{\text{DSF}}^{[0]}=\Vert \delta n^{[0]} \Vert$ (open circles),
$\Delta_{\text{DSF}}^{[k_0]}=\Vert \delta n^{[k_0]} \Vert$ (solid circles)  and DV indicators
$\Delta_{\text{DV}}^{\{\ell\},\text{OEP}}= \Vert
n^{\{\ell\}}-n^{\text{OEP}} \Vert$ (solid squares),
$\Delta_{\text{DV}}^{\{\ell\},\{\ell-1\}}= \Vert
n^{\{\ell\}}-n^{\{\ell-1\}} \Vert$ (open squares) for consecutive
approximations $n^{\{\ell\}}$ to the exchange-only OEP electron
density $n^{\text{OEP}}$ in the Ne atom. The results are obtained
for the approximate exchange potentials $v_{\text{x}}^{[k]}$  found
--- for each $n^{\{\ell\}}$ --- with the algorithm of  Sec.\ \ref{S6}
using $k_0=2 $ steps in the iteration of $v_{\text{x}}^{[k]}$ and
employing (a) the $(N-1)$-parameter optimization  of the ES, and (b)
the $N$-parameter optimization. The starting density $n^{\{0\}}$  is
found with the self-consistent KLI exchange potential
$v_{\text{x}}^{\text{KLI}}$. The dotted line follows the path of the
the $v_{\text{x}}$ iteration.
 }
\label{fig-Ne_dsf_dv_k0=2}
\end{figure}

Further  calculations are performed
using the iterations described in Sec. \ref{S6},
within the exchange-only KS scheme.
The convergence of the  iterated density $n=n^{\{\ell\}}$
is monitored using  the DV indicator
$\Delta_{\text{DV}}^{\{\ell\},\{\ell-1\}}= \Vert
n^{\{\ell\}}-n^{\{\ell-1\}} \Vert$.
The discrepancy  of $n^{\{\ell\}}$ from the exact OEP density  $n^{\text{OEP}}$
is measured  with another DV indicator
$\Delta_{\text{DV}}^{\{\ell\},\text{OEP}}= \Vert
n^{\{\ell\}}-n^{\text{OEP}} \Vert$.
It is found in our test for Ne
(Fig.\ \ref{fig-Ne_dsf_dv_k0=2}) that both  indicators
$\Delta_{\text{DV}}^{\{\ell\},\{\ell-1\}}$,
$\Delta_{\text{DV}}^{\{\ell\},\text{OEP}}$  decrease
rapidly with $\ell$ (like $\ell^{-\zeta}$
where $\zeta$ -- constant) and they are of the same order
at  each step $\ell$ of the density iteration
up to $\ell=\ell_1=20$ where
$\Delta_{\text{DV}}^{\{\ell\},\text{OEP}}$ starts to saturate at
a very small value of the order of $10^{-7}$.
Therefore, a further decrease of
$\Delta_{\text{DV}}^{\{\ell\},\{\ell-1\}}$
for $l>l_1$, seen in Fig.\ \ref{fig-Ne_dsf_dv_k0=2},
and indicating a seemingly better convergence of
$n^{\{\ell\}}$,  is not accompanied by  further improving
of the density (lowering of
 $\Delta_{\text{DV}}^{\{\ell\},\text{OEP}}$) for $l>l_1$.
Similar $\ell$-dependence of the DV indicators
is found for other closed-shell atoms.
Thus, we conclude that
the DV indicator $\Delta_{\text{DV}}^{\{\ell\},\{\ell-1\}}$
is a suitable tool to assess the quality of
the consecutive approximations $n^{\{\ell\}}$
to $n^{\text{OEP}}$, provided that $\ell \leq \ell_1$.
In practice, the value of $\ell_1$
is close to the crossing point of
the $\Delta_{\text{DV}}^{\{\ell\},\{\ell-1\}}$ vs $\ell$ line
with the  $\Delta_{\text{DSF}}^{[0]}$ vs $\ell$ line; see
discussion below.
With this definition, we find
$\ell_1=21,\,20,\,17$ for Ne, Ar, Zn.

In the reported test for the Ne atom,
the exchange potential is calculated using only
$k_0=2$ iteration steps for each approximate density $n=n^{\{\ell\}}$.
Despite so small $k_0$,
the DSF indicator $\Delta_{\text{DSF}}^{[0]}$ decays quickly,
like $\ell^{-\zeta_1}$ (where $\zeta_1$ --- constant)
with increasing $\ell$ during the density iteration
and it approaches a limiting value, less than $10^{-7}$, after
some number of steps close to $\ell_1$ (but, a few steps earlier
for other atoms, like Ar and Zn).
The partner indicator  $\Delta_{\text{DSF}}^{[k_0]}$
reaches the same limiting level in the same range of $\ell$.

The speed of  convergence of iterated density
does not depend on the type
of the ES optimization  used in calculation of
$v_{\text{x}}$, as it is seen in Fig.\ \ref{fig-Ne_dsf_dv_k0=2}(a),(b)
for Ne and in Table \ref{tab-dv-Zn} for Zn.
It is also found that increasing the number $k_0$ of steps in the iteration
of the exchange potential ---
although leading to  smaller $\Delta_{\text{DSF}}^{[k_0]}$
for each density $n^{\{\ell\}}$
%$v_{\text{x}}^{[k_0]}[n^{\{\ell\}}]$ closer
%to $v_{\text{x}}^{\text{E}}[n^{\{\ell\}}]$
---
does not accelerates the overall convergence  in the density iteration.
%This is clearly seen in Table \ref{tab-dv-Zn} where
%Indeed,
In the test for Zn (Table \ref{tab-dv-Zn}),
the  DV indicator
$\Delta_{\text{DV}}^{\{\ell\},\text{OEP}}$ obtained
for $k_0 \geq 3$
is  even slightly larger, than for $k_0=2$;
it is true for each $\ell >0$ and
both types of the ES optimization.
Thus, $k_0=2$ and $(N-1)$-parameter ES modification
is recommended as an optimal choice.

Summarizing the discussion on the use of the  DSF and DV indicators
we conclude that
the condition (\ref{eq-rel-DSF[k_0]}), with suitably chosen
accuracy thresholds $\eta_{\text{DSF}}$
and $\eta_{\text{DV}}$, can indeed be used
for termination of
the combined iteration of $n^{\{\ell\}}$ and $v_{\text{x}}^{[k]}$.
In particular,
in the discussed test for Ne (with $k_0=2$ and
the $(N-1)$-parameter ES optimization)
choosing
$\eta_{\text{DSF}}=\eta_{\text{DV}}=
10^{-4},\,10^{-5},\,10^{-6},\,10^{-7}$
leads to termination after, respectively,
$\ell=6,\,10,\,15,\,20$ steps of the density iteration while
the resultant $n^{\{\ell\}}$ becomes increasingly accurate
since we find
$\Delta_{\text{DV}}^{\{\ell\},\text{OEP}} \approx
\Delta_{\text{DV}}^{\{\ell\},\{\ell-1\}}
\approx \eta_{\text{DV}}$.

The occupied orbital energies
$\epsilon_{nl}$, obtained with the KS potential
$v_{\text{s}}^{\{\ell\}}$
at consecutive iteration steps $\ell$,
converge to the corresponding OEP values $\epsilon_{nl}^{\text{OEP}}$
at similar speed as $\Delta_{\text{DV}}^{\{\ell\},\text{OEP}}$.
For the Ne atom (Fig. \ref{fig-Ne_eorb}), the discrepancies
$|\epsilon_{nl}-\epsilon_{nl}^{\text{OEP}}|$
 decrease with $\ell$ roughly as  $\ell^{-\zeta}$
(i.e., with similar exponential decay as
$\Delta_{\text{DV}}^{\{\ell\},\text{OEP}}$) and saturate
at around the $10^{-6}-10^{-7}\,\text{Hartree}$ level
after a number of steps close to $\ell_1=21$.
Note that $\epsilon_{nl}$ may oscillate around
$\epsilon_{nl}^{\text{OEP}}$ during iteration, which leads to
changing the sign of $\epsilon_{nl}-\epsilon_{nl}^{\text{OEP}}$, like
for $(nl)=2p$ orbital in the Ne atom.

The discrepancy of $\epsilon_{nl}$ from $\epsilon_{nl}^{\text{OEP}}$
becomes less than $10^{-4}\,\text{Hartree}$
for all occupied orbitals $nl$ after 9 density iterations
for Ne, and 12 steps for Ar and Zn while
using only $k_0=2$ steps in iteration of $v_{\text{x}}$ for each density.
Thus, the 0.1 mHartree  accuracy of orbital energies is obtained
after a similar number of iterations as  in Ref.\ \onlinecite{KP03},
where a different method of $v_{\text{x}}$ iteration is applied  and
the OS are found by solving the KP differential equations.

The accuracy of calculated energies $\epsilon_{nl}$
depends on the choice of $p$ and $\kappa$
in the FD formulas (\ref{eq-der-numES}), (\ref{eq-der-numOS})
used to determine the approximate ES and OS
applied in the iteration of $v_{\text{x}}$.
Indeed, in the test for Zn (Table \ref{tab-eorb-Zn}),
the discrepancy of $\epsilon_{nl}$ from $\epsilon_{nl}^{\text{OEP}}$
decreases rapidly with $\kappa$, especially for  $p=3,\,4$.
The discrepancy magnitudes depend on $\kappa$ roughly as  $\kappa^p$
so that they obey the same power law as
the ES and OS errors corresponding to
the type of the FD formula applied in the calculations.
The accuracy of
the occupied orbital energies obtained  for Zn is
$10^{-5}$ and $ 10^{-6}$ $\text{Hartree}$
for $p=2, \kappa=0.05$ and  $p=4$, $\kappa=0.1$,
while it reaches  $10^{-7}\,\text{Hartree}$
 for $p=4$, $\kappa=0.05$.
 Note that the choice of $p$, $\kappa$, though determining the final
 accuracy  of $\epsilon_{nl}$,
 does {\em not} affect  the rate of convergence of the orbital energies
 during density iteration.

\section{Conclusions}

The performed calculations for closed-(sub)shell atoms
prove that the exact exchange potential can be determined very accurately
with the presently proposed algorithm using only the occupied solutions
of the perturbed KS  equations.
Since the respective perturbation is given by the (scaled) DFT exchange
potential or the (scaled) difference between  this  potential and
the non-local Fock exchange operator (built with the KS orbitals),
the perturbed KS equations
have a similar form as the KS or HF equations.
Thus, these KS-like and HF-like equations can be solved with
(only slightly modified) standard numerical methods applicable to % used for
the KS and HF equations.
Therefore, the present algorithm for  determination of
the exact exchange potential,
shown in this work to be effective and very accurate in the
spatial-grid representation (with the pseudospectral method),
is plausibly suitable for its implementation within standard codes using
an orbital basis to represent the KS orbitals as well as
the KS-like and HF-like orbitals applied in the iteration of $v_{\text{x}}$.

The applied method of iterating the exchange potential, which is
the second new element of our algorithm, is found to be convergent
{\em only if} the modification of its OS term, performed at even iteration steps,
is accompanied by the optimization of the ES term at odd steps.
With both ES and OS modifications included, the rapid convergence of
 the obtained DS norm,  decreasing below  $10^{-6}$
in less than 20 iteration steps, proves high efficiency of
the algorithm and  very high accuracy of $v_{\text{x}}$ at convergence
(as the DS norm vanishes for exact $v_{\text{x}}$
due to the OEP equation).
Since  the iteration of the exchange potential can be reduced
to only 2 steps (for each  density)
in the self-consistent density calculations within the exchange-only
KS scheme, without affecting the quick convergence
of the iterated density,
the orbital energies of $10^{-6}-10^{-7}\,\text{Hartree}$ accuracy are obtained
for atoms with only the total number of $20 \times 2 =40$ iterations of $v_{\text{x}}$
(starting with the KLI approximation).

Finally, we expect that the proposed algorithm for
determination of the exact exchange potential,
successfully tested  in calculations for atoms
in a spatial-grid representation, should also be
equally efficient for molecules, and that it could be applied
within an orbital-basis representation.

\begin{acknowledgments}
This work was supported by the Ministry of Science and Higher
Education (grant No. N~N204\ 275939).
\end{acknowledgments}

\appendix
%######################################
\section{Reduction of $N$-dimensional algebraic problem \protect\\
to $(N-1)$-dimensional one}
\label{SA1}
%@@@@@@@@@@@@@@@@@@@@@@@@@@
The $N$-dimensional system of algebraic equations for unknown  ${\bf
{\sf d}}^{'}$, represented by a matrix equation
\begin{equation}
{\bf {\sf A}}^{'}\,{\bf {\sf d}}^{'} = {\bf {\sf b}}^{'}
\label{eq-A1-1}
\end{equation}
where ${\bf {\sf A}}^{'}$ is $N\times N$ square symmetric matrix
$\{A_{ac}\}_{a,c=0}^{N-1}$, while ${\bf {\sf d}}^{'}$ and ${\bf {\sf
b }}^{'}$ are $N\times 1$ column matrices $\{d_c\}_{c=0}^{N-1}$ and
$\{b_c\}_{c=0}^{N-1}$, can be rewritten equivalently as a system of
the following two equations
\begin{eqnarray}
A_{00}\,d_0 + {\bf {\sf A}}_0^{\text{{\sf T}}}\,{\bf {\sf d}} &=&
b_0\,,
\label{eq-A1-2}\\
{\bf {\sf A}}\,{\bf {\sf d}} + {\bf {\sf A}}_0\,d_0 &=& {\bf {\sf
b}}\,, \label{eq-A1-3}
\end{eqnarray}
where ${\bf {\sf A}}$ is $(N-1)\times (N-1)$ square matrix
$\{A_{ac}\}_{a,c=1}^{N-1}$, ${\bf {\sf d}}$, ${\bf {\sf b }}$ and
${\bf {\sf A}}_0$ are $(N-1)\times 1$ column matrices
$\{d_c\}_{c=1}^{N-1}$, $\{b_c\}_{c=1}^{N-1}$,
$\{A_{c0}\}_{c=1}^{N-1}$, while  $A_{00}$, $d_0$ and $b_0$ are
separate elements. The solution of the system of Eqs.\
(\ref{eq-A1-2}) and (\ref{eq-A1-3}) is found to be
\begin{equation}
d_0 = \frac{b_0-{\bf {\sf A}}_0^{\text{{\sf T}}}\,{\bf {\sf
A}}^{-1}\,{\bf {\sf b}} } {A_{00} - {\bf {\sf A}}_0^{\text{{\sf
T}}}\,{\bf {\sf A}}^{-1}\,{\bf {\sf A}}_0}\,,\label{eq-A1-4}
\end{equation}
\begin{equation}
{\bf {\sf d}} = {\bf {\sf A}}^{-1}\,({\bf {\sf b}} - {\bf {\sf
A}}_0\,d_0)\,. \label{eq-A1-5}
\end{equation}
As we see, for its evaluation it is sufficient to determine the
$(N-1)\times (N-1)$ matrix ${\bf {\sf A}}^{-1}$ instead of the
$N\times N$ matrix $\bigl({\bf {\sf A}}^{'}\bigr)^{-1}$, necessary
in the case of direct solution of the original Eq.\ (\ref{eq-A1-1}).
And what is more, in our application the matrix ${\bf {\sf A}}^{-1}$
is common for all steps of iterations at fixed density, while the
matrix $\bigl({\bf {\sf A}}^{'}\bigr)^{-1}$ is specific to each step
separately.
%

%\pagebreak

\widetext
\newcolumntype{e}[1]{D{.}{.}{#1}}

\begin{table}
\caption{ \label{tab-ES-Ne} The approximate ES
$\delta\epsilon_{nl}=\,^{[p,\kappa]}\overset\circ\epsilon_{nl}$ obtained
with the FD formulas  (\ref{eq-der-numES}) for various $p\,,\kappa$.
The calculations use the exchange-only OEP density  $n^{\text{OEP}}$
in  the Ne atom and the corresponding exact exchange potential
$v_{\text{x}}^{\text{E}}=v_{\text{x}}^{\text{E}}[n^{\text{OEP}}]$,
both determined  with the
non-iterative method of Ref. \onlinecite{CH07}.
The shown values are discrepancies of
$\delta\epsilon_{nl}$  from the ES
$D_{nl,nl}^{\text{OEP}}=D_{nl,nl}[v_{\text{x}}^{\text{E}}]$ (given
in the last row). }

\begin{ruledtabular}
\begin{tabular}{dde{7}e{7}e{7}} % dddd e{9}de{9}e{9}e{9}
  &   &
  \multicolumn{3}{c}{$\delta\epsilon_{nl}-D_{nl,nl}^{\text{OEP}}$
  $\;\;(10^{-4}\text{Hartree})$}  \\[2mm]
\multicolumn{1}{l}{$p$} & \multicolumn{1}{c}{$\kappa$} &
\multicolumn{1}{c}{$\;\;(nl)\,=\;1s$} & \multicolumn{1}{c}{$\;\;\;2s$} &
 \multicolumn{1}{c}{$\;\;2p$} \\[2mm]
\hline \\[2mm]
   % after \\: \hline or \cline{col1-col2} \cline{col3-col4} ...

1  &  0.2    &   -0.811010  &  -0.182120   &    -0.245732  \\
1  &  0.1    &   -0.403044  &  -0.090118   &    -0.122396  \\
1  &  0.05  &   -0.200911  &  -0.044839   &    -0.061091  \\ [2mm]

 2  &   0.2   &  -0.009635    &   -0.003484   &    -0.001826 \\
 2   &  0.1   &  -0.002408    &   -0.000879   &    -0.000458 \\
 2   &  0.05 &  -0.000601    &   -0.000220   &    -0.000116 \\ [2mm]

3  &  0.2    &   0.000282  &  0.000435   &   0.000086    \\
3  &  0.1    &   0.000035  &  0.000049   &   0.000008   \\
3  &  0.05  &   0.000006  &  0.0000002 &  -0.000006   \\ [2mm]

4  &  0.2    &   0.000008    &   0.000054   &     0.000010  \\
4  &  0.1    &   0.0000003  &   0.000 004  &    -0.000002 \\
4  &  0.05  &   0.000001  &  -0.000 004   &   -0.000002 \\ [2mm]

\hline \\
\multicolumn{2}{c}{$D_{nl.nl}^{\text{OEP}}$}&
\multicolumn{1}{c}{$1.950\,526\,191\,9 $}  &
\multicolumn{1}{c}{$0.210\,906\,681\,7$}  &
\multicolumn{1}{c}{$\;0$}  \\
\multicolumn{2}{c}{ $\;\;(\text{Hartree})$} \\
\end{tabular}

\end{ruledtabular}

\end{table}

\begin{table*}
\caption{ \label{tab-dv-Zn} The indicator
$\Delta_{\text{DV}}^{\{\ell\},\text{OEP}}=\Vert
n^{\{\ell\}}-n^{\text{OEP}} \Vert $ of the discrepancy between the
approximate electron density $n^{\{\ell\}}$ at the $\ell$-th
iteration step and the exchange-only OEP density $n^{\text{OEP}}$
for the Zn atom. The  shown results are obtained for each density
$n^{\{\ell\}}$ with $k_0$ steps in
the iteration of  the exchange potential $v_{\text{x}}^{[k]}$ and
either $(N-1)$- or $N$-parameter ES optimization. }
 \begin{ruledtabular}
\begin{tabular}{ccccccc}
  & \multicolumn{6}{c}{$\Delta_{\text{DV}}^{\{\ell\},\text{OEP}}$ $\;\;(\times 10^{-3})$}  \\
  & \multicolumn{3}{c}{$(N-1)$-parameter}  & \multicolumn{3}{c}{$N$-parameter}  \\[1mm]
 $\ell $  & $k_0=2$  & $k_0=3$    & $k_0=7$  & $k_0=2$ & $k_0=3$    & $k_0=7$ \\[2mm]
% $\ell $ & \multicolumn{8}{c}{$(\times 10^{-3})$ } \\[2mm]
\hline \\[2mm]
  % after \\: \hline or \cline{col1-col2} \cline{col3-col4} ...

0  &     3.07390  &   3.07390    &   3.07390  &   3.07390  &   3.07390    &   3.07390 \\
1  &     1.27749  &   1.71563    &   1.82028  &   1.27749  &   1.84049   &   1.83820  \\
2  &     0.95047  &   1.06142    &   1.07960  &   0.79210  &   1.09074    &   1.08965  \\
3  &     0.49044  &   0.61554    &   0.64348  &   0.46845  &   0.65009    &   0.64967  \\
5  &     0.17837  &   0.21939    &   0.22823  &   0.16534  &   0.23049    &   0.23052  \\
10  &   0.01355  &   0.01659    &   0.01724  &   0.01242  &   0.01739   &   0.01742  \\
20  &   0.00049  &   0.00060    &   0.00060  &   0.00049  &   0.00060   &   0.00060  \\
30  &   0.00047  &   0.00057    &   0.00057  &   0.00047  &   0.00057    &   0.00057
\end{tabular}
\end{ruledtabular}
\end{table*}

\begin{table*}
\caption{ \label{tab-eorb-Zn} The orbital energies  $\epsilon_{nl}$
vs.\ $p\,,\kappa$
for the Zn atom obtained --- after 25 iterations of  $n^{\{\ell\}}$
--- with the algorithm of Sec. \ref{S6} using $k_0=2$ iteration steps
and $(N-1)$-parameter ES optimization in  determination of the
exchange potential  $v_{\text{x}}^{[k]}$ for each $n^{\{\ell\}}$.
The applied OS and ES are found with the FD formulas
(\ref{eq-der-numES}), (\ref{eq-der-numOS}) using the chosen $p\,,
\kappa$. The shown values are discrepancies of $\epsilon_{nl}$ from
the orbital energies $\epsilon_{nl}^{\text{OEP}}$ (given in the last
row) calculated  with the exact exchange potential
$v_{\text{x}}^{\text{E}}[n^{\text{OEP}}]$ determined  with the non-iterative method
of Ref. \onlinecite{CH07}. }

 \begin{ruledtabular}
\begin{tabular}{ce{-2}e{-4}e{-4}e{-4}e{-4}e{-4}e{-4}e{-4}} % e{-4}e{-4}e{-4}e{-4}e{-4}e{-4}e{-4}e{-4}
  &   &
  \multicolumn{7}{c}{$\epsilon_{nl}-\epsilon_{nl}^{\text{OEP}}$
  $\;\;(10^{-4}\text{Hartree})$}  \\[2mm]
  %
  % &  \multicolumn{1}{c}{$\Delta_{\text{DV}}^{\{25\},\{24\}} $}  &
  % \multicolumn{1}{c}{$\Delta_{\text{DSF}}^{[k_0]}[n^{\{25\}}]$} \\
  %
   % \multicolumn{9}{c}{  } &$(\times 10^{-9})$ & $ (\times 10^{-6})$   \\[2mm]

 $p$ & \multicolumn{1}{c}{$\kappa$} & \multicolumn{1}{c}{$(nl)\,=\;\;1s$} &
 \multicolumn{1}{c}{$2s$} &  \multicolumn{1}{c}{$3s$} &
 \multicolumn{1}{c}{$4s$} &
  \multicolumn{1}{c}{$2p$} &
   \multicolumn{1}{c}{$3p$} &
 \multicolumn{1}{c}{$3d$} \\[2mm]
\hline \\[2mm]
  % after \\: \hline or \cline{col1-col2} \cline{col3-col4} ...

  1 & 0.2 &
    -10.640  &  -9.992   &  -12.802   &   3.106 &
    -8.231 &   -12.603  &  -10.352   \\
   1 & 0.1 &
  -5.192   &  -4.714   &  -6.001   &    1.474  &
  -3.813   &  -5.889   &   -4.808  \\
  1 & 0.05 &
  -2.575   &  -2.296   &  -2.912  &   0.719 &
  -1.842   &  -2.853  & -2.323   \\[2mm]

 2 & 0.2 &
 -0.056   &   -1.075 &   -1.885  &   0.308  &
 -0.981   &   -1.957 &    -1.816  \\
 2 & 0.1 &
 -0.019 &    -0.268 &    -0.466 &   0.077 &
    -0.246   &  -0.484 &   -0.449  \\
 2 &  0.05      &
-0.005  & -0.066 & -0.116  &   0.019 &
-0.061  & -0.121  & -0.112   \\[2mm]

3 & 0.2 &
0.126 &    0.661&    0.927 &   -0.131 &
0.640  &  0.963 & 0.914  \\
3 & 0.1 &
 0.010  &  0.065 & 0.091 & -0.013&
0.062 & 0.095 & 0.090\\
3 & 0.05 &
0.001  & 0.009   & 0.011  & -0.002 &
 0.007 &  0.011 & 0.010  \\[2mm]

4 & 0.2 &
 0.041 &  0.170  & 0.225 &   -0.031 &
0.165 &  0.233& 0.222 \\
4 & 0.1 &
 0.002  &   0.011  &    0.013 &  -0.002 &
 0.010 &    0.013 &    0.013 \\
 4 & 0.05 &
-0.0005 &   0.0015 &  0.0007 &  -0.0004 &
 0.0006 &   0.0005 &  0.0007  \\[2mm]

\hline \\
% &   &
%  \multicolumn{6}{c}{$\epsilon_{nl}^{\text{OEP}}$   $\;\;(\text{Hartree})$} \\[2mm]
% $p$ & $\kappa$ & $(nl)\,=\;\;1s$ & $2s$ & $3s$ & $4s$ & $2p$ & $3p$ & $3d$ \\[2mm]
%\multicolumn{2}{c}{\text{exact}} &
\multicolumn{2}{c}{$\epsilon_{nl}^{\text{OEP}}$}&
\multicolumn{1}{c}{$-345.755720523$} &
\multicolumn{1}{c}{$-41.714189169$} &
\multicolumn{1}{c}{$ -4.796168733$} &
\multicolumn{1}{c}{$-0.292805644$} &
\multicolumn{1}{c}{$-36.742098912$} &
\multicolumn{1}{c}{$-3.210661901$} &
\multicolumn{1}{c}{$-0.537803838$} \\
\multicolumn{2}{c}{ $\;\;(\text{Hartree})$} \\
\end{tabular}

\end{ruledtabular}

\end{table*}

\begin{figure}
\includegraphics*[width=9cm]{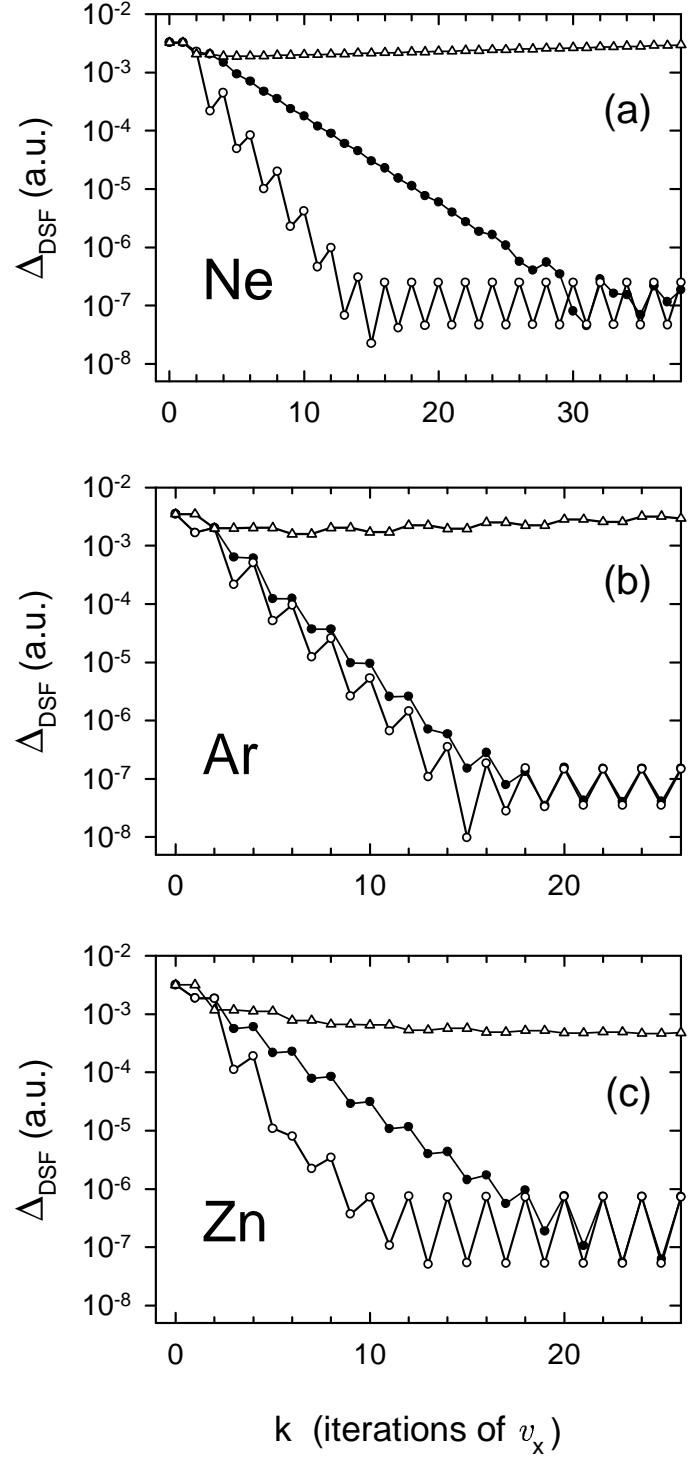}
\caption{ DSF indicator $\Delta_{\text{DSF}}^{[k]}=\Vert \delta
n^{[k]} \Vert$  vs.\ the index  $k$ of the iterative
approximation $v_{\text{x}}^{[k]}$ to the exact exchange potential
for the exchange-only OEP density $n^{\text{OEP}}$ in  (a) Ne, (b) Ar
and (c) Zn atoms. The respective $v_{\text{x}}^{[k]}$ is
determined with the algorithm of  Sec.\ \ref{S6}, for {\em even} $k$
--- by modifying the OS term in  $v_{\text{x}}^{[k-1]}$, for {\em
odd} $k$  ---  by the ES term modification: the $(N-1)$-parameter
optimization  of the ES (solid circles), the $N$-parameter
optimization  (open circles), no optimization, i.e., empty step
(open triangles).
 }
\label{fig-Ne_dsf_noep}
\end{figure}

%
%\FloatBarrier
%

\end{document}